\documentclass[article]{elsarticle}

\usepackage{setspace}
\usepackage{lineno}
\usepackage{hyperref}
\usepackage[hyphenbreaks]{breakurl}
\usepackage{subcaption}
\usepackage{pdflscape}
\usepackage{xspace}

\usepackage{color,soul}
\usepackage[dvipsnames]{xcolor}
\usepackage{siunitx}
\usepackage{algorithm}
\usepackage[noend]{algpseudocode}
\usepackage{mathtools}
\usepackage{amsfonts}
\usepackage{bibunits}

\def\colb{\textcolor{black}}
\def\colr{\textcolor{red}}
\def\colr{\textcolor{black}}

\newcommand{\beginsupplement}{%
        \setcounter{table}{0}
        \renewcommand{\thetable}{S\arabic{table}}%
        \setcounter{figure}{0}
        \renewcommand{\thefigure}{S\arabic{figure}}%
     }

\begin{document}

\begin{frontmatter}

  \title{An Efficient Monte Carlo Algorithm for Determining the
    Minimum Energy Structures of Metallic Grain Boundaries}
  
    % My title suggestion:
    %    1) A Monte Carlo Algorithm for the Efficient Sampling of the Grain Boundary Energy Landscape
    % Please add your suggestions here:
    % 
  \author[mymainaddress]{Arash Dehghan Banadaki}
  \author[marksaddress]{Mark A. Tschopp}
  \author[mymainaddress]{Srikanth Patala\corref{mycorrespondingauthor}}
  \cortext[mycorrespondingauthor]{Corresponding author}
  \ead{spatala@ncsu.edu}

  \address[mymainaddress]{Department of Materials Science and
    Engineering, North Carolina State University, Raleigh, NC, 27617, USA}

  \address[marksaddress]{U.S. Army Research Laboratory, Aberdeen
    Proving Ground, MD 21005, USA}

  \begin{abstract}
    Sampling minimum energy grain boundary (GB) structures in the
    five-dimensional crystallographic phase space can provide
    much-needed insight into how GB crystallography affects various
    interfacial properties. However, the complexity and number of
    parameters involved often limits the extent of this exploration to
    a small set of interfaces. In this article, we present a fast
    Monte Carlo scheme for generating zero-Kelvin, low energy GB
    structures in the five-dimensional crystallographic phase
    space. The Monte Carlo trial moves include removal and insertion
    of atoms in the GB region, which create a diverse set of GB
    configurations and result in a rapid convergence to the low energy
    structure. We have validated the robustness of this approach by
    simulating over 1184 tilt, twist, and mixed character GBs in both
    fcc (Aluminum and Nickel) and bcc ($\alpha$-Iron) metallic
    systems.

  \end{abstract}

%% \begin{abstract}
%%   Sampling minimum energy grain boundary (GB) structures in the
%%   five-dimensional crystallographic phase space can provide
%%   much-needed insight into how GB crystallography affects various
%%   interfacial properties. However, the complexity and number of
%%   parameters involved often limits the extent of this exploration to a
%%   small set of GBs. In this article, we present a fast Monte Carlo
%%   scheme, which include the removal and insertion of GB atoms, for
%%   generating zero-Kelvin, minimum energy GB structures. We have
%%   validated the robustness of this approach by simulating over 1184
%%   tilt, twist, and mixed character GBs in both face-centered and
%%   body-centered cubic metallic systems.
%% \end{abstract}

  % \begin{keyword}
  %   \CBred{Grain Boundaries \sep \sep Energy \sep Monte Carlo \sep Crystallography}
  % \end{keyword}
\end{frontmatter}

% \linenumbers

\begin{bibunit}[unsrt]

%%%%%%%%%%%%%%%%%%%%%%%%%%%%%%%%%%%%%%%%%%%%%%%%%%%%%%%%%%%%%%%%%%%%%%%%%%%%%%%%%%%%

%%%%%%%%%%%%%%%%%%%%% \input{Introduction}

\section*{Introduction}
Grain boundaries (GBs) influence a wide array of mechanical
\cite{lehockey1997creep, lehockey1998improving, chen2000role,
bechtle2009grain, kobayashi2010grain, gertsman2001study,
king2008observations}, chemical \cite{mishin1999grain,
chen2007percolation, fujita2004using} and functional
\cite{babcock1995nature, frary2003nonrandom} properties in
polycrystalline materials.  However, they are also among the least
understood defect types due to the vast and topologically complex
five-dimensional (5-D) crystallographic phase-space of interfaces
\cite{patala2013symmetries, patala2012improved}. In other words, the
GB properties are functions of, at the least, five macroscopic
crystallographic degrees of freedom (DOF).

In general, for single component metallic systems, there exist
\emph{nine} crystallographic parameters that uniquely define the
structure of a GB \cite{kalonji1982symmetry}. These parameters are
classified into macroscopic and microscopic DOF---five parameters
specifying the misorientation and boundary plane orientation and the
additional four representing the microscopic relative displacements
between the adjoining lattices and the translation of the boundary
plane along its normal vector. However, under conditions of
thermodynamic equilibrium, it is generally accepted that the five
macroscopic DOF are sufficient for representing the properties of
interfaces \cite{lejcek2010grain}.

For developing reliable GB structure-property relationships, the
lowest energy GB structures, computed at \SI{0}{\kelvin}, are
essential.  The energy landscape and the structure of a GB with fixed
crystallography (the five macroscopic parameters) depends on the
microscopic DOF and the atomic density $\lambda$
\cite{hickman2017extra} (traditionally controlled by tuning the
allowed extent of overlap between atoms). In the past few decades,
several, reasonably successful efforts have been made to predict the
low-energy GB structures
\cite{rittner1996simulations,tschopp2007structures,tschopp2007structural,
  olmsted2009survey, erwin2012continuously, banadaki2016simple}.
While the implementation varies slightly, these techniques generally
rely on generating a large number of initial GB configurations by
varying the microscopic DOF of an interface, which can be considered
as a brute-force approach for determining the minima in the energy
landscape. While such a brute-force approach might suffice for
simulating GBs with low $\Sigma$-number \cite{grimmer1976coincidence},
the computational cost usually increases as the symmetry of the GB is
reduced.

Monte Carlo (MC) based algorithms have been routinely utilized for
finding minima in energy landscapes in a variety of complex systems in
condensed-matter physics \cite{landau2014guide}.  However, such a
technique has never been applied to computing the minimum energy
structures for GBs in single component systems. This is primarily due
to the fact that atoms along the interface are not constrained to lie
on a fixed lattice \cite{landau2014guide}.  For example, hybrid
Monte-Carlo/Molecular-Dynamics simulations have been utilized to
compute low energy GB structures in binary alloy systems
\cite{pan2016effect, pan2017formation}. These alloys have at least two
components and the trial moves correspond to swapping the positions of
\emph{unlike} atoms. Unfortunately, in single component systems, atom
swapping does not change the configuration of the bicrystal. In this
article, we introduce a Monte Carlo based GB energy minimization
algorithm applicable for single component systems. The advantage of a
MC framework is that, when the acceptance probabilities are devised
appropriately \cite{newman1999monte}, it can be utilized to compute
thermodynamic equilibrium properties in a variety of relevant
statistical ensembles.

As will be discussed in the next section, the trial perturbations that
facilitate the MC-based approach involve both atom removal and, more
importantly, atom insertions in the GB region. These two Monte-Carlo
moves change the density of the GBs and are likely the most important
perturbations for sampling the energy landscape of the microscopic DOF
of an interface. There have been recent studies where atom removal and
insertions were utilized to compute minimum energy structures or to
investigate phase transitions in GB %grain boundary
structures. For example, in \cite{frolov2013structural}, the atomic
density was allowed to vary by removing periodic boundary conditions
in the GB plane. The free surfaces act as sources and sinks for
atoms. This facilitated the required changes in grain boundary density
and a structural phase transition was observed. In
\cite{zhu2018predicting} and \cite{frolov2018grain}, low energy GB
structures were obtained using a genetic algorithm where, among
others, atom removal and insertion were used to perturb the GB
structure. The insertions were made by constructing a uniform grid in
the GB region and filling the unoccupied grid points at random.

For a Monte-Carlo simulation to work in an efficient manner, it is
important that the increase in energy due to the perturbations are not
always too large. However, if atoms are inserted randomly, the
Monte-Carlo simulations will take a long time to converge due to low
acceptance rates. In this article, we introduce a geometric
construction to identify voids for atom insertion in the GB region,
which alleviates the large increases in energy. This technique is
similar to the cavity based MC method developed for the simulation of
dense fluids \cite{mezei1980cavity}. Inserting atoms in these voids
and minimizing the structure dramatically improves the acceptance
probability of the atom-insertion move. As far as the authors are
aware, there has been only one previous report that utilized atom
insertions and removals in a Monte-Carlo framework for defects in
crystalline systems. Phillpot and Rickman \cite{phillpot1992simulated}
proposed a grand canonical framework, where \emph{sites} with
fractional occupancies instead of atoms are used, to obtain ground
state structures in the presence of defects. In
\cite{phillpot1992simulated}, a simple Lennard-Jones potential is used
to compute the minimum energy structure of $(110)$ twist GB. However,
in this technique, the sites for insertion are determined a priori and
are fixed during the MC simulation. Our algorithm builds on these
ideas and shows that low-energy configurations for GBs can be obtained
for a diverse set of GB crystallographic characters in both fcc
(Aluminum and Nickel) and bcc ($\alpha$-Iron) metallic systems.

% \colb{Adding or removing atoms has to be done in integral amounts. For
%   a gas or a dilute liquid this presents little problem for simulation
%   because the removal or addition of a single atom has relatively
%   little effect on the total energy of the system. Indeed, there have
%   been a number of grand-canonical Monte Carlo (GCMC) simulations of
%   gases and dilute liquids \cite{}. As the density of system
%   increases, however, due to the presence of nearby atoms, the initial
%   energy change on adding an additional atom increases, and
%   consequently, the efficiency of GCMC simulations decreases
%   dramatically. In particular, attempts to insert an atom into dense
%   systems usually fail. as there are not many voids into which the atom
%   can go. The }

The MC approach, introduced in this article, is also very efficient in
generating the low energy GB structures. The biggest obstacle for
generating large GB databases that are well-sampled in the 5-D
crystallographic phase-space is the massive number of simulations
required to obtain the lowest-energy GB structure. For example,
determining the lowest energy structure for a typical GB using such
brute force algorithms requires anywhere between \num{1000} to
\num{150000} unique energy-minimization simulations. In this article,
we also show that the proposed Monte Carlo scheme is more efficient in
generating the low-energy GB structures when compared to such
traditional brute-force simulations. In the following sections, we
describe the Monte Carlo algorithm, the trial moves and the test
cases, involving the three GB databases, in greater detail.

\section*{Methodology}

Our Monte Carlo algorithm starts with an initial GB configuration and
applies random perturbations (trial moves), which are then evaluated
using a Metropolis-like criterion \cite{allen1989computer}: accepted
if the energy is reduced, accepted or rejected by a Boltzmann-weighted
probability if the energy increases. At each step of the simulation,
the following perturbations may be introduced: (a) removal of an atom
from the GB and (b) insertion of an atom in the GB region.

The trial move involving \emph{atom removal} (or creation of a vacancy) is
inspired by investigations of von Alfthan et
al.~\cite{von2006structures}, Yu and Demkowicz \cite{yu2015non}, and
Tschopp et al.~\cite{tschopp2014binding}. Initially, we only
considered trial moves involving atom removal and realized that
applying just this perturbation does not result in the low-energy
structure in many of the test cases (described in the later part of
the article). We observed that an efficient convergence to the low
energy GB structure is obtained by also considering perturbations that
involve \emph{atom insertions} at GB interstitial sites.
% As such insertion of atoms is 

\begin{figure}[h!]
  \centering
  \includegraphics[width=\linewidth]{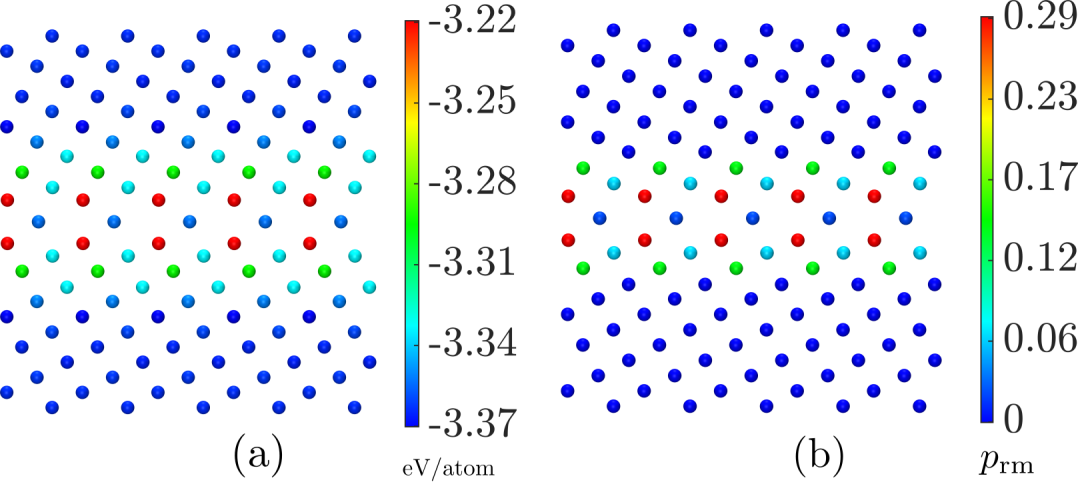}
  \caption{Illustration of removal probabilities. (a) The atomistic
      structure of $\Sigma 5 (0 \, \bar{2}\, 1)$ GB, where the atoms
      are colored according to their energies.  (b) The same
      $\Sigma 5 (0 \, \bar{2}\, 1)$ GB structure is shown with the
      atoms colored according to their removal probabilities.}
  \label{fig:rsites}
\end{figure}

The algorithm starts with an initial random GB configuration, which is
created using a random set of microscopic DOF for the interface. The
next step involves choosing one of the two trial moves stochastically
(i.e., the removal and insertion moves are chosen with a probability
of $p_{\text{rm}}$ and $p_{\text{in}} = 1 - p_{\text{rm}}$,
respectively).  \colb{Once the decision for removal or insertion has
  been made, the} atom to remove or the interstitial site for atom
insertion is also chosen stochastically to facilitate the possibility
of generating a diverse set of GB structures. For example, to
determine the GB atom to remove \footnote{In FCC and BCC bicrystals,
  the centrosymmetry parameter \cite{kelchner1998dislocation}, as
  computed by LAMMPS, is used to identify GB atoms (with the criterion
  of CSP $ > \, 0.1$).}, we first assign a removal probability
$p_{rm, i}$, for each atom $i$ in the GB, which is given by:

\begin{equation} 
\label{eq:rmprob} 
p_{\text{rm}, i} = 
\begin{cases}
    \left( E_i -    E_{\circ} \right) / \left( \sum_{j=1}^{N_{\text{GB}}}(E_j - E_{\circ}) \right),& \text{if } E_i \geq E_{\circ} \\        0,         & \text{otherwise}
  \end{cases}
\end{equation}

\noindent where $E_i$ is the energy of the $i^{\text{th}}$ GB atom,
$E_{\circ}$ is the cohesive energy of the atom in the single crystal
configuration at \SI{0}{\kelvin}, $N_{\text{GB}}$ is the number of GB
atoms, and \colb{$\sum p_{\text{rm}, i} = 1$}. According to this
equation, the probability of removing an atom is proportional to its
excess energy. \colb{In principle, we are interested in removing an
  atom that lowers the GB energy. Such an atom can be determined by
  computing the minimum GB vacancy formation energy defined in
  \cite{yu2015non}. However, to determine the atom that corresponds to
  the minimum vacancy formation energy, one would have to remove a GB
  atom, relax the structure, compute the vacancy formation energy, and
  then repeat this procedure for all the atoms in the GB. This is
  computationally very expensive. Instead, we simply choose to remove
  atoms based on their excess energies. The excess energies can be
  directly determined from the GB configuration and no further
  simulations are required. This choice is further motivated by prior
  studies that have shown, for example, that the excess energy is
  positively correlated with the formation energy of certain GB
  defects (e.g., He and He$_2$ in a monovacancy in
  Ref.~\citenum{tschopp2014binding}).} In Figure~\ref{fig:rsites}(a),
the atomistic structure of a $\Sigma 5 (0 \, \bar{2}\, 1)$ GB is
shown, where the atoms are colored according to their energy.
Corresponding removal probabilities for the atoms are shown in
Figure~\ref{fig:rsites}(b).

% proportional to the excess energy (the difference between the atom's
% energy and the cohesive energy of the lattice). While it is more
% likely that high-energy atoms are removed from the GB, there exist
% trial moves, even though with low probability, where atoms with low
% excess energy are removed from the GB. This allows for a more diverse
% sampling of the GB structures during the Monte Carlo simulation and,
% hence, increases the chance of obtaining the lowest-energy
% configuration.

Similarly, the sites for atom insertion are also chosen stochastically
within the GB. The importance of the insertion step has been
highlighted in a Cu-Al binary alloy system
\cite{campbell2004copper}, where the copper atoms preferentially segregate
to the interstitial sites in the $\Sigma 5 (3 \, 1 \, 0)$ Aluminum
GB. This result underscores the importance of considering atom
insertion steps during the Monte-Carlo simulation for achieving faster
energy convergence.

\begin{figure}[h!]
  \centering
  \includegraphics[width=\linewidth]{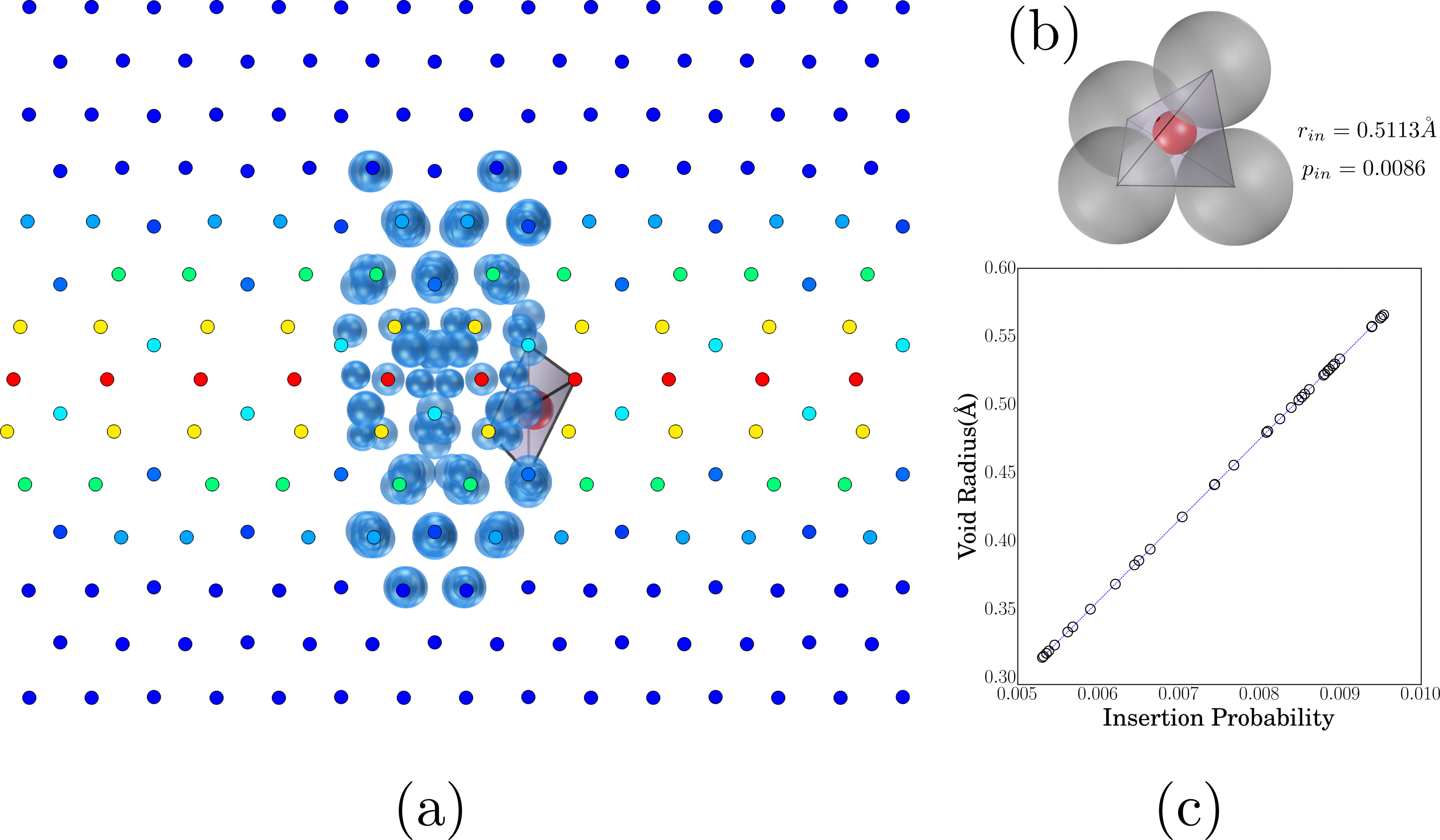}
  \caption{Illustration of insertion probabilities.  (a) The
      atomistic structure of an asymmetric tilt GB,
      $\Sigma 3 (1 \, 0 \, \bar{1})$, is shown along with the voids
      present in the GB structure. The voids are obtained by meshing
      the GB structure and computing the circumcenters of the Delaunay
      tetrahedra. One such tetrahedron is shown in the GB structure
      and its magnified version is illustrated in (b). The radius of
      the void is given by the difference between the radius of the
      circumsphere and the atom radius. The probability of atom
      insertion is directly proportional to the void radius, as shown
      in (c).}
  \label{fig:asites}
\end{figure}

The potential interstitial sites for inserting an atom in the GB
region are determined through a Delaunay triangulation
\cite{lee1980two} of the GB atomistic structure. The circumcenters of
the Delaunay tetrahedra provide the locations of the interstitial
voids.  For example, Figure \ref{fig:asites}(a) shows the voids that
are identified at the circumcenters of the Delaunay tetrahedra of the
$\Sigma 3 (1 \, 0 \, \bar{1})$ GB \footnote{We chose this asymmetric
  tilt GB to simply illustrate the algorithm for computing voids in
  the GB structure. The same concept can be utilized for any complex
  GB structure.}. To simplify presentation, we show only one Delaunay
tetrahedron within the GB. A magnified version of this tetrahedron and
the void is shown in Figure \ref{fig:asites}(b). The radius of the
\colb{interstitial} void is given by \colb{$r_{in} = r_s - r_a$},
where $r_s$ is the radius of the circumsphere of the Delaunay
tetrahedron and $r_a$ is the radius of the atom.

\begin{figure}[h!]
  \centering
  \includegraphics[width=0.85\linewidth]{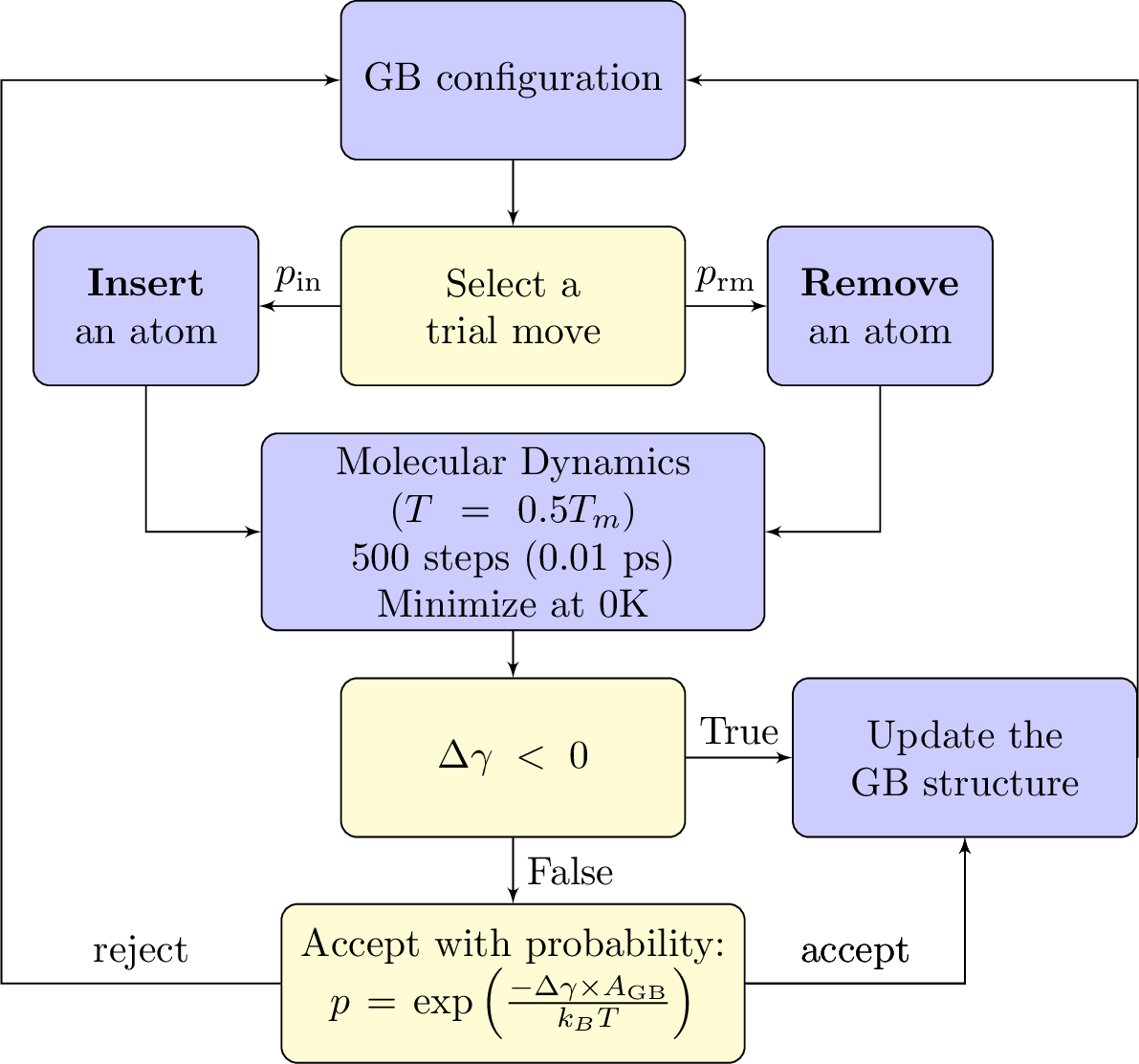}
  \caption{A flowchart of different steps involved in simulating a GB
    using the proposed Monte-Carlo simulation technique.}
  \label{fig:fc}
\end{figure}

In a recent study, we showed that the interstitial voids, determined
using the circumcenters of the Delaunay tetrahedra, accurately predict
the sites of hydrogen segregation in certain Ni GBs (refer to
Figure~S30 in \cite{banadaki2017three}). Once the radii of all the
interstitial voids within the GB region are computed, one of the sites
is chosen stochastically with the probability:

\begin{equation} 
\label{eq:inprob} 
p_{\text{in}, i} =  \frac{r_{v, i}}{\sum_{j=1}^{N_v}(r_{v,j})}
\end{equation}

\noindent where $r_{v,i}$ is the radius of the $i^{th}$ void, $N_v$ is
the total number of GB interstitial sites considered, and
\colb{$\sum p_{\text{in}, i} = 1$}. Figure \ref{fig:asites}(c) shows
the insertion probabilities $p_{in}$ for the voids in
$\Sigma 3 (1 \, 0 \, \bar{1})$ GB, and, as expected, $p_{in}$ is
proportional to $r_{v,i}$. In summary, the trial moves consist of
stochastically choosing either \emph{an atom to remove} or \emph{an
  interstitial site for atom insertion}. After the appropriate trial
move, the new structure is relaxed at \SI{0}{\kelvin} in LAMMPS
\cite{plimpton1995lammps}. \colb{If the difference in the GB energy
  before and after the trial move is denoted by $\Delta \gamma $, the
  new GB configuration is accepted if $\Delta \gamma < 0$. If
  $\Delta \gamma > 0$, the new GB configuration is accepted with a
  probability of
  $\exp \left( \frac{- \Delta \gamma \times A_{\text{GB}}}{k_B T_R}
  \right)$, }where $A_{\text{GB}}$ is the area of the GB in the
simulation box, $k_B$ is the Boltzmann constant and $T_R$ is the
material specific reference temperature. Upon acceptance of a trial
perturbation, the GB energy and configuration are updated and the
perturbations are repeated for the new structure. Otherwise, the
perturbed structure is discarded and the trial moves are repeated. As
a summary, the MC algorithm is illustrated as a flowchart in
Figure~\ref{fig:fc}.

%%%%%%%%%%%%%%%%%%%%%%%%%%%%%%%%%%%%%%%%%%%%%%%%%%%%%%%%%%%%%%%%%%%%%%%%%%%%%%%%%%%%

%%%%%%%%%%%%%%%%%%%%%%%%%%%%%%%%%%%%%%%%%%%%%%%%%%%%%%%%%%%%%%%%%%%%%%%%%%%%%%%%%%%%
%%%%%%%%%%%%%%%%%%%%%%%% \input{Results}
\section*{Results and Discussion}

To test the efficiency of this Monte Carlo scheme, we simulated the
set of 388 GBs in Aluminum and Nickel from Olmsted et
al.~\cite{olmsted2009survey} and the set of 408 GBs in $\alpha$-Iron
from Ratanaphan et al.~\cite{ratanaphan2015grain}. The interatomic
potentials used to simulate these GBs are also identical with the
exception of Aluminum where a more accurate Embedded Atom Method (EAM)
potential developed by Mishin et al.~\cite{mishin1999interatomic} is
utilized. This EAM potential function has been shown to reproduce the
material parameters crucial for interface simulations - the cohesive
energy, lattice parameter and the stable and metastable stacking fault
energies - with reasonable accuracy when compared to density
functional theory computations. For Al, to account for the change in
EAM potential, we repeated the brute force energy calculations for the
388 GBs.

For the Monte Carlo simulations, the initial configuration for each GB
is generated by choosing a random set of the microscopic DOF (i.e.,
the relative bicrystal displacement and boundary-plane translation
along the normal vector). The bicrystallographic aspects, required for
simulating the GBs in LAMMPS, are computed using the the GBpy software
(\href{https://pypi.python.org/pypi/GBpy}{https://pypi.python.org/pypi/GBpy})
developed by the authors \cite{banadaki2015efficient}.

% To further improve the efficiency of the
% simulations, only a single unit cell is simulated for certain GBs.
% Since the GB is periodic in two dimensions, with the periodicity of
% the two-dimensional (2-D) coincidence site lattice
% \cite{banadaki2015efficient}, the box dimensions must be multiples of
% this GB unit-cell vectors.  For certain low-symmetry GBs, where the
% units cell dimensions are large, it suffices to simulate just a single
% unit cell.  For such GBs, we use the GBpy software
% (\colb{\href{https://pypi.python.org/pypi/GBpy}{https://pypi.python.org/pypi/GBpy}})
% for determining the smallest 2-D GB unit cell for the Monte Carlo
% simulations .

We find that, in most cases, the proposed Monte Carlo scheme finds the
lowest-energy structure (as compared to the energies reported in
\cite{olmsted2009survey} and \cite{ratanaphan2015grain}) for GBs in
FCC (Al and Ni) and BCC ($\alpha$-Fe) metallic systems in just
$\mathbf{5000}$ steps for the chosen reference temperature
$T_R = 0.5 T_m$, where $T_m$ is the melting temperature, and a value
of $p_{\text{rm}} = 0.5$. \emph{The \num{5000} steps include both
  accepted and rejected trial moves} in the Monte Carlo scheme. We
tried the values of $T_R = (0.25, 0.5, 0.75)T_m$ and
$p_{\text{rm}} = 0.25, 0.5, 0.75$ while optimizing the MC
algorithm. The parameters finally chosen ($T_R = 0.5 T_m$ and
$p_{\text{rm}} = 0.5$) had the highest rate of energy convergence when
compared to the brute-force algorithm.

% \emph{The 5000 steps are the maximum number of
%   trial perturbations applied to all the GBs}.

% \begin{equation} 
% \label{eq:error} 
% \text{\% Error} =
% \frac{E_{\text{GB}}^{\text{Monte-Carlo}}-E_{\text{GB}}^{\text{Brute
%       Force}}}{E_{\text{GB}}^{\text{Brute Force}}} \times 100
% \end{equation}

\begin{figure}[h!]
  \centering
  \includegraphics[width=\linewidth]{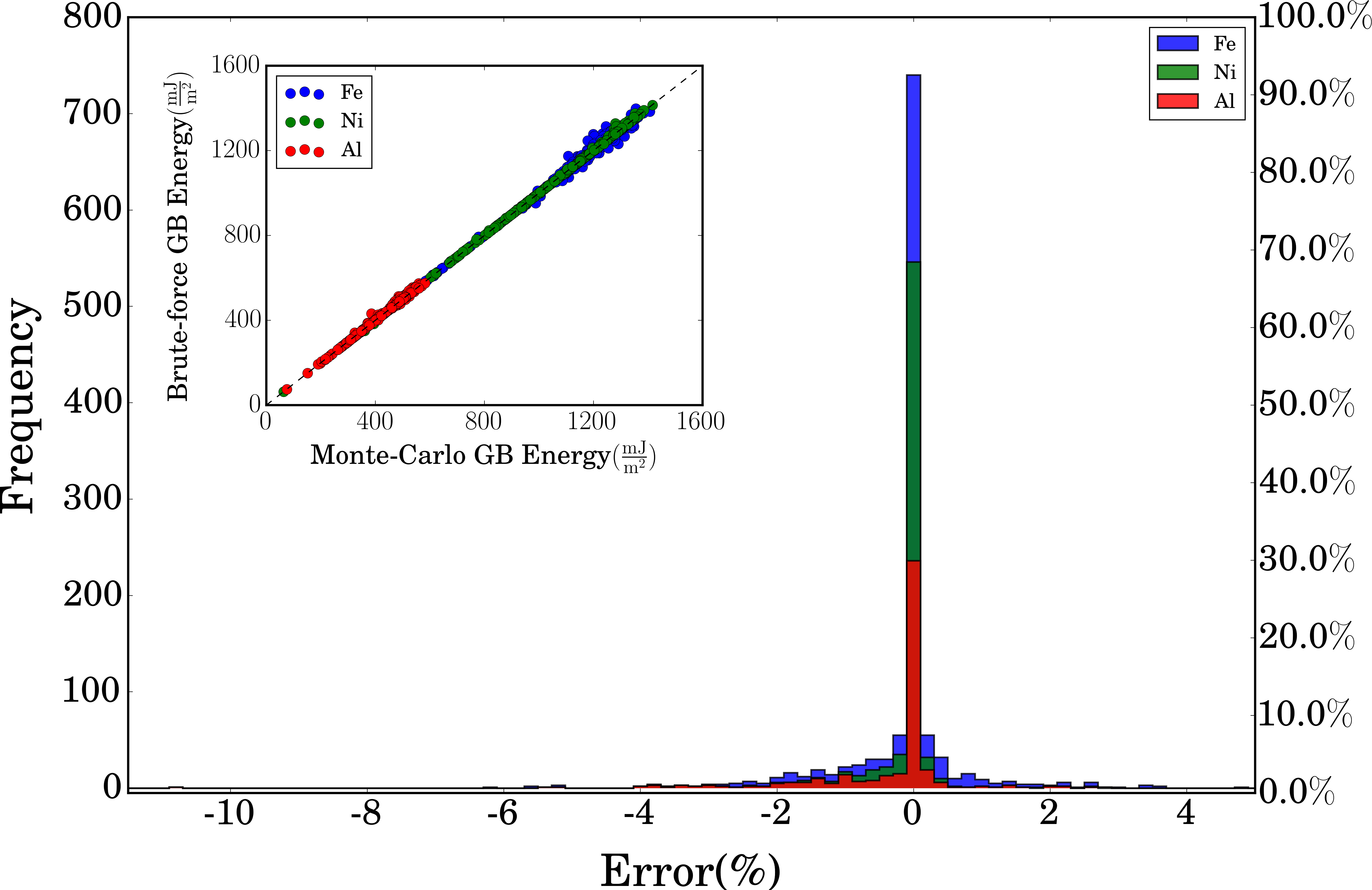}
  \caption{A \emph{stacked} histogram of the Error (\%) for the 388
    GBs of Aluminum, 388 GBs of Nickel and 408 GBs of $\alpha$-Iron is
    shown. A negative value indicates that the energy obtained through
    the Monte Carlo scheme is lower than that obtained using the
    brute-force algorithm (i.e., the MC scheme provides a better
    estimate for the lowest-energy structure of the GB for these
    cases). \colr{Also shown in the inset is a correlation plot
      between the energies obtained through the MC vs. the brute-force
      algorithm.}}
  \label{fig:eng_hist}
\end{figure}

In Figure~\ref{fig:eng_hist}, we compare the performance of the
proposed algorithm by plotting a histogram of the error percentage
defined as:

\begin{equation} 
\label{eq:error} 
\text{\% Error} =
\frac{\gamma^{\text{Monte-Carlo}}-\gamma^{\text{Brute-Force}}}{\gamma^{\text{Brute-Force}}} \times 100
\end{equation}

\noindent That is, \% Error the difference in the energies obtained
from the MC and the brute-force techniques divided by the brute-force
energy. The histogram reports the \% Error for all the 1184 simulated
interfaces (388 Al, 388 Ni and 408 $\alpha$-Fe GBs). This figure
illustrates the accuracy with which the minimum energy GB structures
are realized in the Monte Carlo scheme---approximately 96\% of the
1184 GBs simulated have an error less than 1\% and only about 0.6\% of
the GBs exhibited an error greater than 3\%. The maximum error is
$\sim 4.8\%$ for
$\Sigma 315 (2 \, 1 \, 3) (\bar{1} \, \bar{2} \, \bar{3})$ GB of
$\alpha$-Fe. About 11\% of the GBs show an error less than $-1\%$,
which indicates that, in some cases, the Monte Carlo scheme results in
lower energy structures than the ones obtained through the brute force
simulations. Finally, in Figures~\ref{fig:alfzs}, \ref{fig:nifzs} and
\ref{fig:fefzs}, the \% error is illustrated in the complete 5-D phase
space of the GBs (i.e. as a function of the five macroscopic
crystallographic DOF of GBs). It is evident from these figures that
there are no obvious correlations between GB crystallography and \%
Error.

We also tested the influence of the initial configuration on the
performance of the MC simulations for Aluminum GBs (as we have access
to a diverse set of GB structures from the brute-force
simulations). In addition to the random configuration, we performed
the MC simulations with two additional configurations as the initial
state:

\begin{itemize}
\item \textbf{Maximum-energy}: GB structure corresponding to the
    highest-energy configuration from the brute-force simulations.
\item \textbf{Most-frequent}: GB configuration that occurs with
  highest frequency in the brute-force simulations but \emph{does not
    correspond to the lowest-energy state}. This structure may
  represent a deep local minimum in the energy landscape.
\end{itemize}

\begin{figure}[h!]
  \centering
  \includegraphics[width=1.0\linewidth]{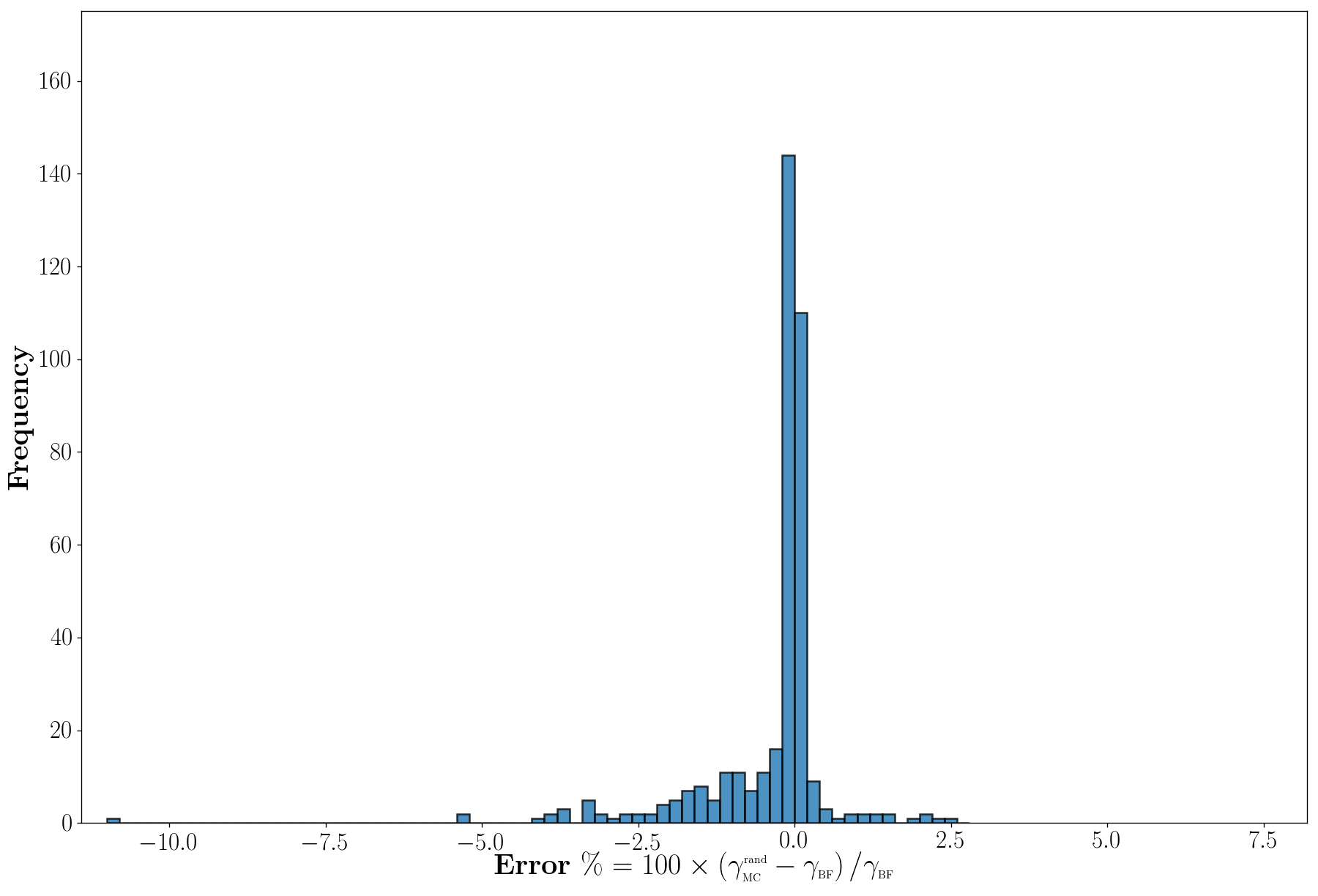}
  \caption{Histogram of the Error (\%) for the 388 GBs of Aluminum is
    shown. The initial configurations for the MC simulations are
    constructed with random microscopic DOF.}
  \label{fig:err1a}
\end{figure}

\begin{figure}[h!]
  \centering
  \includegraphics[width=1.0\linewidth]{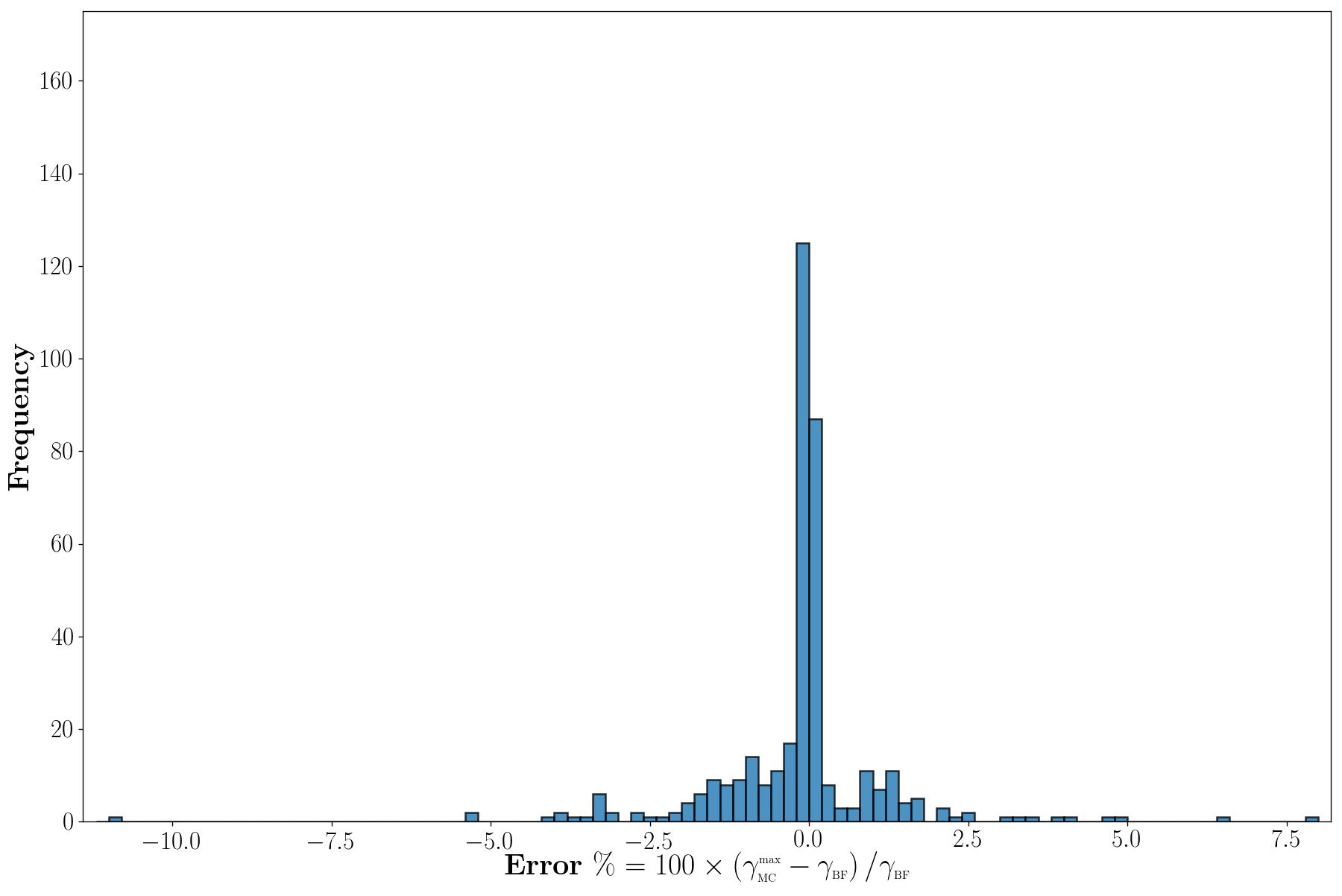}
  \caption{Histogram of the Error (\%) for the 388 GBs of Aluminum is
    shown. The initial configurations are the maximum energy
    structures obtained from the brute-force simulations performed for
    Aluminum GBs.}
  \label{fig:err1b}
\end{figure}

\begin{figure}[p]
  \centering
  \includegraphics[width=1.0\linewidth]{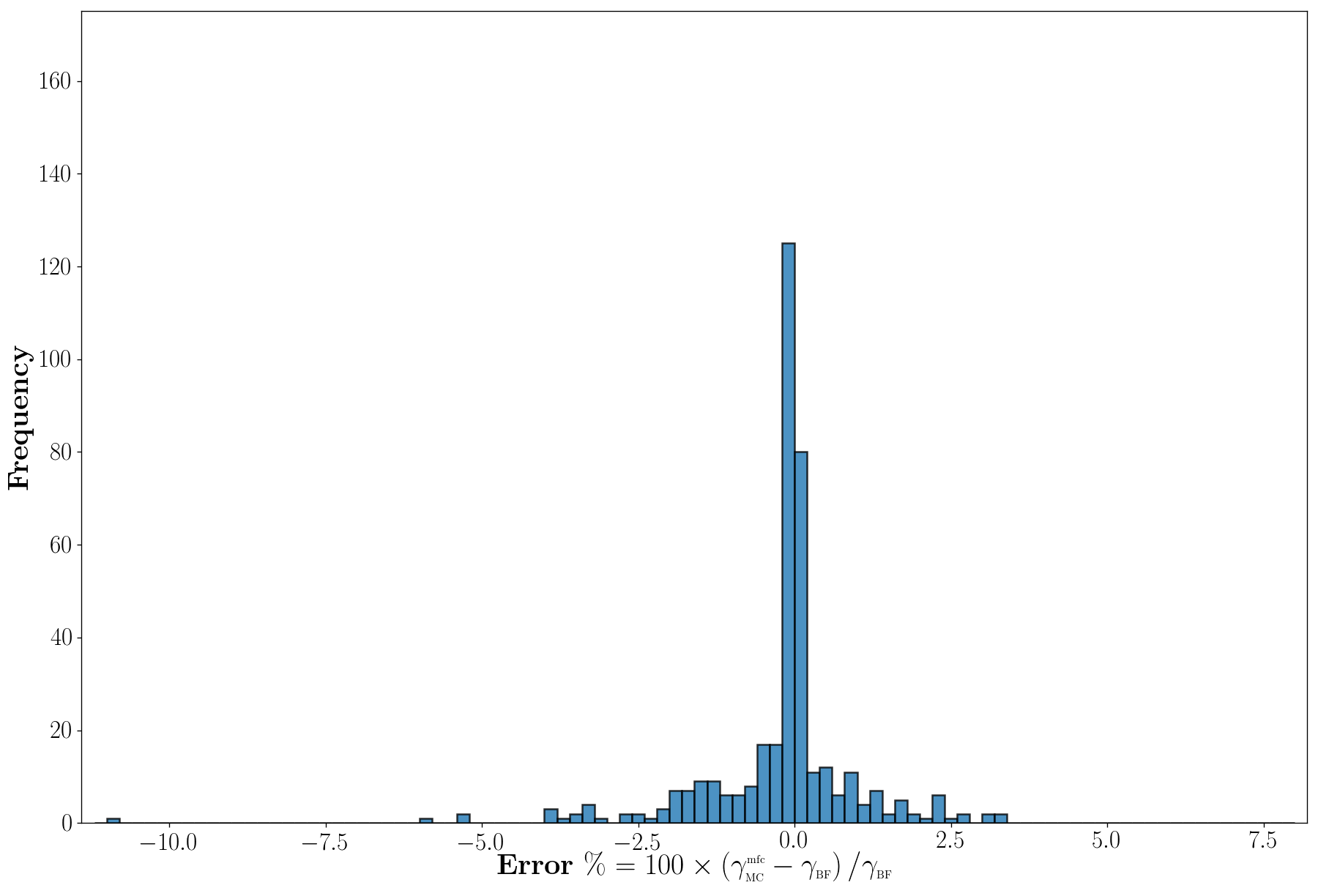}
  \caption{Histogram of the Error (\%) for the 388 GBs of Aluminum is
    shown. The initial configurations correspond the most frequent
    structures (other than the minimum energy structure). These
    structures are obtained from the brute-force simulations performed
    for Aluminum GBs.}
  \label{fig:err1c}
\end{figure}

In Figure \ref{fig:err1a}, we plotted the error percentages for MC simulations
with random initial configurations. In Figures \ref{fig:err1b} and
\ref{fig:err1c}, the error percentages for the MC simulations with initial
configuration as maximum-energy and most-frequent structures are plotted,
respectively. The maximum-energy structure performs poorly compared to the
random or the most-frequent structure. \colb{To compare the performance of these
  simulations with different initial structures, we will use the sum of the
  errors greater than zero. That is, we use the quantity $\sum_{i} \max \left(
  0, \gamma_{\tiny{\mbox{MC}}} - \gamma_{\tiny{\mbox{BF}}} \right)$, where
  $\gamma_{\tiny{\mbox{MC}}}$ and $\gamma_{\tiny{\mbox{BF}}}$ are the GB
  energies obtained using the Monte-Carlo and Brute-Force simulations,
  respectively. For the maximum-energy, the most-frequent, and the random
  initial structures, this quantity results in 109.5 mJ/m$^2$, 89.5 mJ/m$^2$ and
  26.4 mJ/m$^2$, respectively. That is, the simulations where the initial
  structures correspond to the maximum-energy configurations perform the worst,
  but the most-frequent initial structure is not far behind. These results are
  not unexpected given that the simulations with the maximum-energy structure as the
  initial configuration can be interpreted as the worst-case scenario for the MC
  algorithm. For the most-frequent initial structure case, the larger errors may
  be due to the fact that the GB structure is getting stuck in a local minimum.}

\newpage
\clearpage

\newpage
\clearpage

\newpage
\clearpage

\hspace{0pt}
\vfill

\begin{figure}[p]
  \centering
    \includegraphics[width=0.75\linewidth]{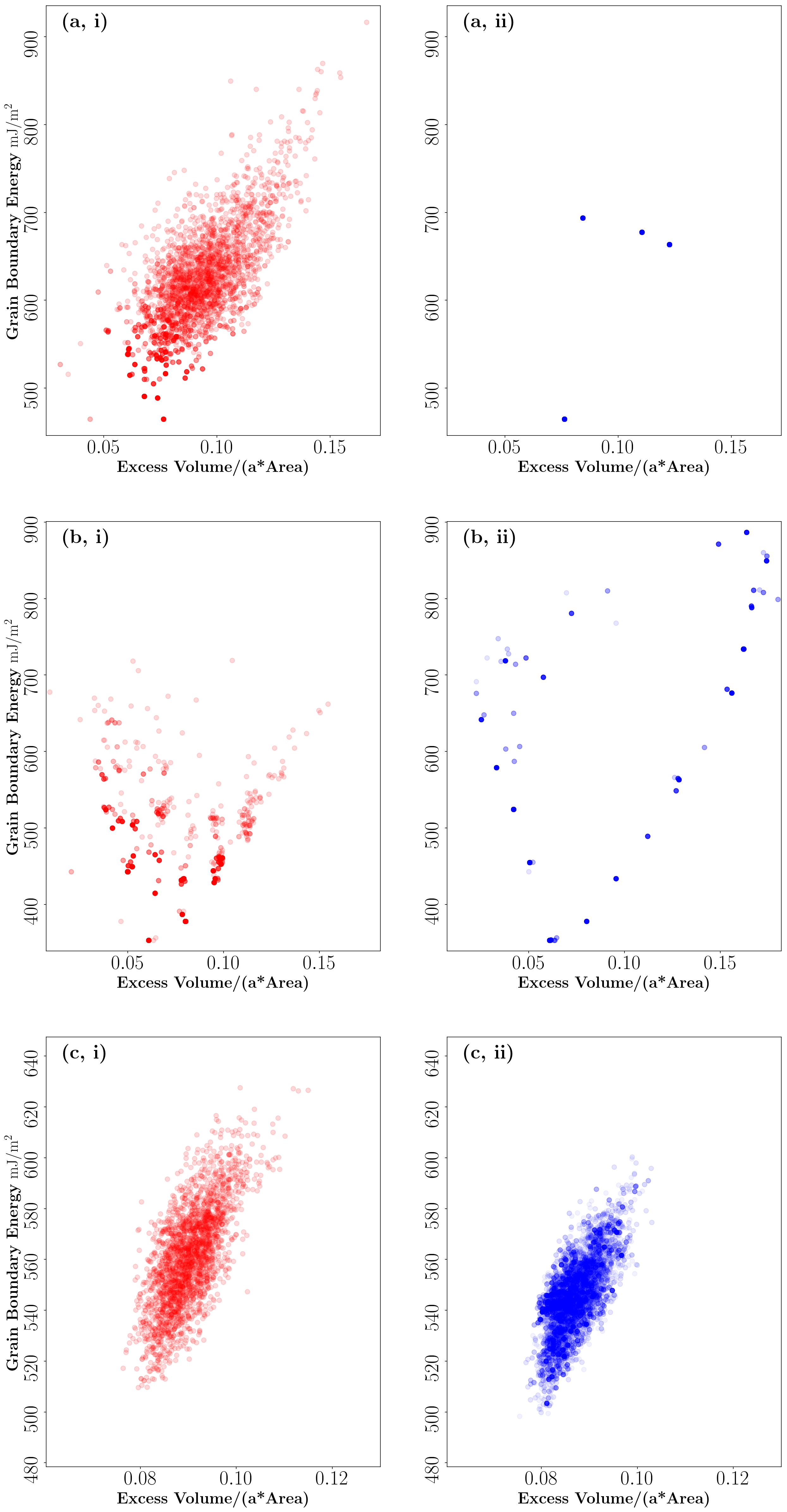}
    \caption{The energies are plotted against the normalized excess volume per
      unit area for three Aluminum GBs: (a) $\Sigma 3 \left( 1 \, 0 \, \bar{1}
      \right)$, (b) $\Sigma 17 \left( 1 \, 0 \, 0 \right)$ and (c) $\Sigma 77
      \left( 9 \, \bar{5} \, 2 \right)$. In (i), the excess volume of the GB
      structures from the Monte-Carlo simulation are plotted. The structures are
      the relaxed structures of the accepted trail perturbations. As a
      comparison, the excess-volume of the relaxed GB structures obtained from the
      brute-force simulations are plotted in (ii).}
  \label{fig:exvol}
\end{figure}

\colb{To better understand the influence of atom insertion and removal on GB
  structures, we provided a comparison of the range of excess-volumes sampled by
  the GB structures obtained through the Monte-Carlo and the brute-force
  simulations. For this analysis, we chose three Aluminum GBs: (a) $\Sigma 3
  \left( 1 \, 0 \, \bar{1} \right)$, (b) $\Sigma 17 \left( 1 \, 0 \, 0 \right)$
  and (c) $\Sigma 77 \left( 9 \, \bar{5} \, 2 \right)$ and, in Figure
  \ref{fig:exvol}, plotted the GB energies vs. the excess-volume per unit area
  (in units of lattice constant). In (i) and (ii), the excess-volumes of the
  Monte-Carlo and brute-force simulations are plotted, respectively. It is
  evident from the scatter in these plots, the diversity of the GB structures
  sampled is much more pronounced than those observed in the brute-force
  simulations.} Therefore, it is reasonable to infer that a diverse set of GB
structures are sampled by simply applying the atom removal and insertion
perturbations. Furthermore, the effect of the trial perturbations on GB energies
is evaluated by plotting the evolution of GB energies during the MC simulation
for a few exemplary cases in section \ref{sec:Smcevol}. The atom removal and
insertion moves are, in some sense, drastic enough that they change both the GB
energies and structures considerably. \colb{While we refer the reader to section
  \ref{sec:Smcevol} for a complete discussion on the trends in the energy
  evolution during the MC simulations, we would like to point out that the MC
  simulations not only provide the lowest-energy GB structure but also sample
  important metastable states \cite{han2016grain}.}

\vfill
\hspace{0pt}

\newpage
\clearpage

\newpage
\clearpage

\hspace{0pt}
\vfill

\begin{figure}[h!]
  \centering
    \includegraphics[width=\linewidth]{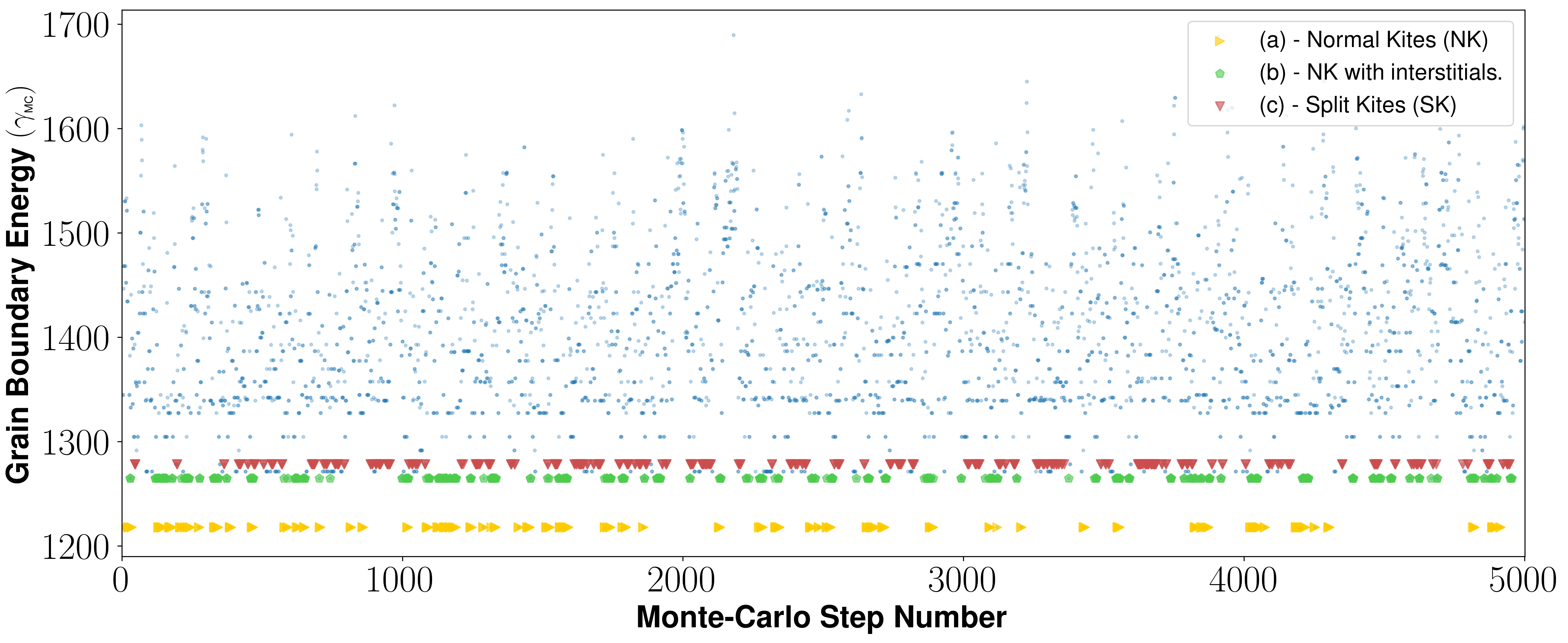}
    \caption{A scatter plot of GB energies vs. the Monte-Carlo step number is
      shown. Highlighted are the three most frequently observed GB structures
      during the Monte Carlo simulation. (a), (b) and (c) were observed 307, 326
      and 388 times, respectively during the simulation. The structures of these
      GBs are shown in Figure \ref{fig:s5gbs}.}
  \label{fig:s5mceng}
\end{figure}

\colb{We use the example of the evolution of structures and energies of $\Sigma
  5 \left( 0 \bar{1} 3 \right)$ GB, in Nickel, to illustrate how the MC
  simulations could shed light on GB metastability. The structural aspects of
  $\Sigma 5 \left( 0 \bar{1} 3 \right)$ GB have been analyzed as function of
  temperature \cite{frolov2013effect, frolov2013structural, frolov2014effect,
    freitas2018free} and, more recently, it has been shown that, in Copper, this
  GB exhibits a phase-transition at approximately 184 $\pm$ 4 Kelvin
  \cite{freitas2018free} from a structure that contains ``normal-kites'' to a
  structure that contains ``split-kites'' \cite{frolov2013effect}. To compute
  the phase-transition temperature using the technique developed in
  \cite{freitas2018free}, both the 0K lowest-energy and the relevant metastable
  structures have to be determined \emph{a priori}. In the MC simulations, GBs
  that are sampled frequently but do not correspond to the lowest-energy
  structure can be considered as relevant metastable structures. For example,
  consider the plot of the energies vs. MC step number for the $\Sigma 5 \left(
  0 \bar{1} 3 \right)$ Nickel GB shown in Figure \ref{fig:s5mceng}. In this
  plot, we highlight three GB structures, labeled (a), (b) and (c), that
  correspond to the top three most sampled structures during the MC simulation.}

\begin{figure}[h!]
  \centering
    \includegraphics[width=\linewidth]{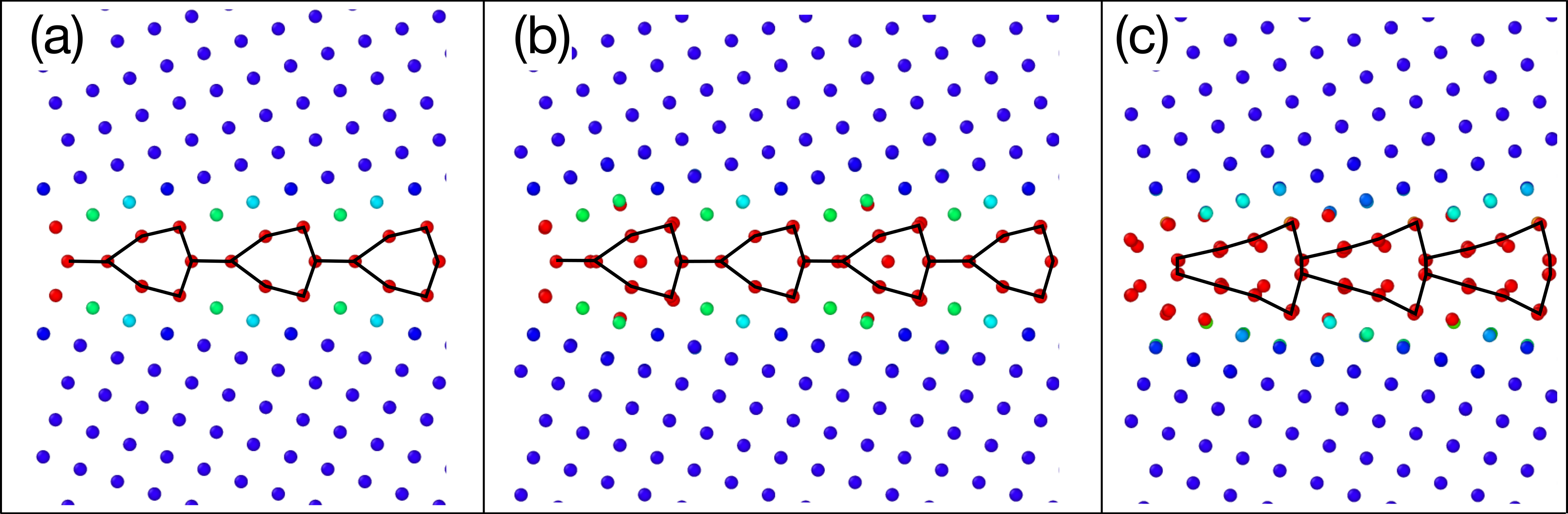}
    \caption{Three structures of $\Sigma 5 \left( 0 \bar{1} 3 \right) $ GBs are
      illustrated in the decreasing order of frequency with which they are
      observed during the Monte-Carlo simulation. (a) The structure with the
      lowest energy exhibits normal-kites and is observed approximately 307
      times of the 5000 Monte-Carlo steps. (b) This structure is observed
      approximately 326 times, has one normal-kite unit and another with an
      interstitial. (c) This GB structure has the split-kite units and has been
      observed approximately 388 times.}
  \label{fig:s5gbs}
\end{figure}

\colb{The lowest-energy GB structure with normal-kite units is shown in Figure
  \ref{fig:s5gbs}(a) and has been observed about 307 times (out of 5000) during
  the Monte-Carlo simulation. However, there are two other structures with
  higher energies that are observed more frequently than the lowest-energy
  structure. These GBs are labeled (b) and (c) in Figure \ref{fig:s5mceng} and
  are shown in Figure \ref{fig:s5gbs} (b) and (c), respectively. The structure
  in Figure \ref{fig:s5gbs}(b) is obtained by simply introducing an interstitial
  in one of the normal-kite units and has been observed about 326 times. The
  structure in Figure \ref{fig:s5gbs}(c) exhibits the split-kite structure, has
  the highest frequency, and has been observed about 388 times during the
  Monte-Carlo simulation. The split-kite structure may very well have been the
  equilibrium structure, for the $\Sigma 5 \left( 0 \bar{1} 3 \right)$ GB, at
  the temperature used in the Boltzmann acceptance probability (T = 0.5T$_m$).
  Therefore, it is important to note that the MC algorithm introduced in this
  article not only generates the lowest-energy structures but also provides the
  most relevant metastable structures. Once these metastable structures are
  determined, techniques such as thermodynamic integration
  \cite{frolov2015segregation, freitas2018free} could be used to determine the
  phase transitions of GBs.}

\vfill
\hspace{0pt}

\newpage

% It is instructive to visualize the evolution of the GB energy during
% the MC simulation.  In Figures \ref{fig:mcevol1}, \ref{fig:mcevol2},
% \ref{fig:mcevol3} and \ref{fig:mcevol4}}

Finally, as mentioned in the introductory paragraphs of the article, the MC
algorithm is computationally more efficient than the brute-force technique in
generating low-energy GB structures. To give an indication of the improved
computational efficiency, Table \ref{table:S1} provides the amount of time
required to generate low-energy structures for 15 GBs, ordered according to
computational expense, with differing bicrystallography. Since the number of
simulation steps required in the brute-force algorithm depends on the
bicrystallographic symmetry, the relative performance also depends on the GB
type. In this table, explicit comparisons for the time required to generate the
minimum energy structures for 15 GBs, with computation times varying from
minimum to maximum, are shown.

\begin{table}[h!]
  \caption{The efficiency of the Monte-Carlo simulation, as compared
    to the brute-force algorithm, is illustrated by tabulating the
    time (in mins) taken by each algorithm to generate the low
    energy GB structure. Both the brute-force and the Monte-Carlo
    simulations for each GB were performed on a single compute node
    with 16 cores, and 64 GB memory.}
  \centering
    \begin{tabular}{ | p{4.0cm} | p{2.0cm} | p{2.0cm} | p{2.5cm} |}
    \hline
GB Crystallography & Brute-Force (mins) & Monte-Carlo (mins) & Relative Performance \\ \hline
$\Sigma 3 (1\,1\,1)(\bar{1}\,\bar{1}\,\bar{1})$		       &	813	        &	166	  &	4.89	\\	\hline
$\Sigma 27 (1\,\bar{1}\,5)(1\,\bar{1}\,\bar{5})$		       &	1875	&	119	  &	15.76	\\	\hline
$\Sigma 101 (1\,0\,0)(\bar{1}\,0\,0)$		                       &	5563	&	319	  &	17.44	\\	\hline
$\Sigma 29(0\,\bar{2}\,5)(0\,\bar{2}\,\bar{5})$		       &	1438	&	154	  &	9.33	\\	\hline
$\Sigma 105(3\,1\,5)(\bar{1}\,\bar{3}\,\bar{5})$		       &	2188	&	140	  &	15.63	\\	\hline
$\Sigma 63(\bar{1}\,10\,\bar{5})(\bar{1}\,\bar{5}\,10)$	       &	1875	&	200	  &	9.37	\\	\hline
$\Sigma 111(11\,\bar{10}\,\bar{1})(\bar{10}\,11\,\bar{1})$   &	2750	&	1526  &	1.80	        \\	\hline
$\Sigma 57(3\,\bar{5}\,2)(\bar{2}\,5\,\bar{3})$		       &	4438	&	158	  &	28.09	\\	\hline
$\Sigma 93(5\,\bar{6}\,1)(\bar{1}\,6\,\bar{5})$		       &	4625	&	197	  &	23.48	\\	\hline
$\Sigma 111(1\,1\,1)(\bar{1}\,\bar{1}\,\bar{1})$		       &	4719	&	165	  &	28.60	\\	\hline
$\Sigma 55(\bar{3}\,\bar{1}\,0)(3\,1\,0)$		               &	9563	&	158	  &	60.52	\\	\hline
$\Sigma 99(\bar{4}\,1\,1)(4\,\bar{1}\,1)$		               &	9844	&	166	  &	59.30	\\	\hline
$\Sigma 95(0\,\bar{1}\,3)(1\,0\,\bar{3})$		               &	5125	&	193	  &	26.55	\\	\hline
$\Sigma 77(\bar{9}\,5\,2)(9\,\bar{5}\,2)$		               &	4438	&	287	  &	15.46	\\	\hline
$\Sigma 109(1\,0\,0)(\bar{1}\,0\,0)$		                       &	5156	&	325	  &	15.87	\\	\hline
    \end{tabular}
\label{table:S1}
\end{table}

On average the Monte Carlo simulations (for 5000 trial moves) were
about 20 times more efficient. One reason for the efficiency is the
fewer number of minimization steps required in the MC scheme. The
other is related to the local nature of the MC perturbations. The
computational time, to minimize a structure in the MC scheme, is a lot
less than that required to minimize the bicrystal configurations
created in brute force algorithms.

\begin{figure}[h!]
  \centering
    \includegraphics[width=1.0\linewidth]{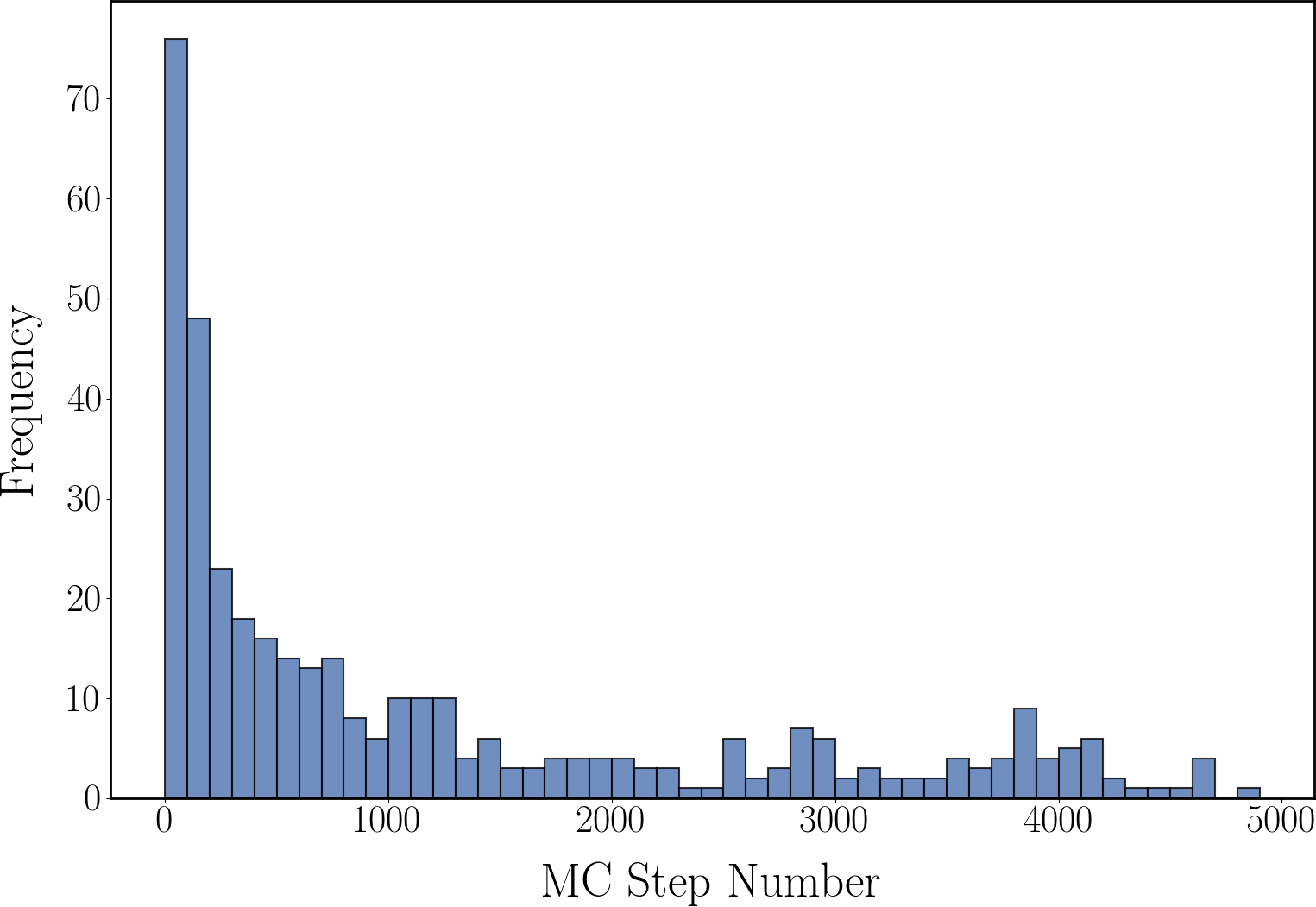}
    \caption{The frequency of the Monte Carlo step numbers when the
      minimum energy structure is encountered for the 388 GBs in
      Aluminum. The initial structure is the maximum energy GB from
      the brute-force simulations.}
  \label{fig:err0}
\end{figure}

It is important to note that the 5000 steps correspond to the total
energy minimizations, including accepted and rejected moves, performed
for each GB. Therefore, 5000 is not an ensemble average of many GBs
but it is the maximum number of steps required to get within 5\% of
the brute-force energy for all 1184 GBs. In the plot below (Figure
\ref{fig:err0}), we show a histogram of the step number (corresponding
to the first Monte Carlo step when the minimum GB energy was obtained)
for all the 388 Aluminum grain boundaries. The initial structure for
the GBs is the maximum energy structure from the brute-force
simulations. It is clear from this plot that for most of the GBs, the
minimum energy structure is recovered within the first 2000-3000
steps.

% For example, we noticed that the simulation of
% $\Sigma 111(11\,\bar{10}\,\bar{1})$, $(\bar{10}\,11\,\bar{1})$,
% $\Sigma 93(5\,\bar{6}\,1)$ and $(\bar{1}\,6\,\bar{5})$ and
% $\Sigma 55(\bar{3}\,\bar{1}\,0)$ $(3\,1\,0)$ GBs were approximately
% 1.8, 23.4 and 60.5 times faster than the brute-force simulations.

%%%%%%%%%%%%%%%%%%%%%%%%%%%%%%%%%%%%%%%%%%%%%%%%%%%%%%%%%%%%%%%%%%%%%%%%%%%%%%%%%%%%

%%%%%%%%%%%%%%%%%%%%%%%%%%%%%%%%%%%%%%%%%%%%%%%%%%%%%%%%%%%%%%%%%%%%%%%%%%%%%%%%%%%%
%%%%%%%%%%%%%%%%%%%%%%%% \input{Conclusions}
\section*{Conclusions}

In conclusion, we present a novel and efficient technique for
generating the low-energy \SI{0}{\kelvin} structures for both FCC and
BCC metallic GBs. It is observed that the Monte Carlo algorithm
developed in this article is on average an order of magnitude faster
than the traditional brute-force technique. Since the number of
simulations in a brute-force algorithm depends on the symmetry aspects
of the interface, the efficiency achieved by the MC simulation is
shown for fifteen exemplary cases in Table \ref{table:S1}. While it is
well recognized that it is necessary to change the number of atoms in
an interface to access the different energetic states, prior
techniques have relied primarily on removal \cite{von2006structures},
swapping of atoms \cite{sadigh2012scalable, pan2016effect}, or on the
insertion of atoms randomly in the GB structure
\cite{zhu2018predicting, frolov2018grain}. The novel aspect of our
algorithm is the geometrically-informed atom insertion as a
perturbation step, which results in an efficient convergence to lower
energy states of the GB.

More recently, the importance of analyzing metastable GB states has
been highlighted by Han et al. \cite{han2016grain}. The proposed MC
algorithm will also help sample these meta-stable states as the trial
perturbations visit a large and diverse range of GB
structures. Therefore, this algorithm not only allows for an efficient
simulation of low-energy GB structures, but also provides a means to
compute thermodynamic equilibrium properties. To this end, we are
currently extending the MC framework to a hybrid
Monte-Carlo/Molecular-Dynamics scheme \cite{sadigh2012scalable}.
Compared to the genetic algorithms that perform grand-canonical based
optimization of GB structures \cite{chua2010genetic,
  zhu2018predicting, frolov2018grain}, the Monte-Carlo technique has
the advantage that the acceptance probabilities can be tailored to
satisfy detailed balance equations for different ensembles. For
example, in \cite{hudson2006grand}, detailed balance equations, for
performing grand canonical Monte Carlo simulations, have been
developed for inter-granular glassy films. Similar equations can be
extended to the GBs and thermodynamic equilibrium properties, such as
the free energy as a function of temperature, can be computed in a
variety of relevant statistical ensembles.
  
%
% \colb{In comparison to the the other grand-canonical simulations
% using genetic algorithms \cite{}, our MC algorithm has a temperature
% scale inherent in the acceptance probabilities. This }

This MC simulation scheme is general enough and can be easily extended
to multi-component systems. We anticipate that adding the
atom-insertion step to the traditional hybrid
Monte-Carlo/Molecular-Dynamics algorithm \cite{sadigh2012scalable},
which only contains the atom swapping events, will result in a faster
convergence, particularly when the size difference between the solute
and solvent atoms is large.

% Even though recent theories suggest the
% importance of analyzing metastable GB states , the
% lowest-energy structures are still crucial for interfacial property
% calculations. Since, under equilibrium conditions, the probability
% $p_i$ of observing a GB state with energy $E_i$ is proportional to
% $\exp(-E_i/k_BT)$, the lowest energy structure will dominate the
% equilibrium property computations (with a weight, $p_i$). 

% Therefore, this algorithm not only allows for an efficient computation
% of the 5-D energy landscape, but also provides the lowest-energy GB
% structures, which are valuable for computing any interfacial property
% (mobility, diffusivity, stiffness, etc.) using atomistic simulations.

%%%%%%%%%%%%%%%%%%%%%%%%%%%%%%%%%%%%%%%%%%%%%%%%%%%%%%%%%%%%%%%%%%%%%%%%%%%%%%%%%%%%

\section*{Acknowledgments}

This work is supported by the AFOSR Young Investigator Program funded
through the Aerospace Materials for Extreme Environments (Contract \#
FA9550-17-1-0145). The computational support was provided by the High
Performance Computing Center at North Carolina State University.

\newpage

\section*{Data availability}

The raw/processed data required to reproduce these findings cannot be
shared at this time as the data also forms part of an ongoing study.

\section*{References}
\putbib[main]
\end{bibunit}

\newpage

\clearpage

\newpage

\beginsupplement
%%%%%%%%%%%%%%%%%%%%% \include{supp_info}
%%%%%%%%%% Merge with supplemental materials %%%%%%%%%%
% \begin{widetext}
\begin{doublespacing}
\begin{center}
  \textbf{ \LARGE Supplemental Materials: \\ An Efficient Monte-Carlo
     Algorithm for Extracting the Minimum Energy Structure of Metals}
\end{center}
\end{doublespacing}

%%%%%%%%%% Merge with supplemental materials %%%%%%%%%%
%%%%%%%%%% Prefix a "S" to all equations, figures, tables and reset the counter %%%%%%%%%%
\setcounter{equation}{0}
\setcounter{figure}{0}
\setcounter{table}{0}
\setcounter{page}{1}
\setcounter{section}{0}

\makeatletter

\renewcommand{\theequation}{S\arabic{equation}}
\renewcommand{\thesection}{S\arabic{section}}
\renewcommand{\thefigure}{S\arabic{figure}}
\renewcommand{\bibnumfmt}[1]{[S#1]}
\renewcommand{\citenumfont}[1]{S#1}
%%%%%%%%%% Prefix a "S" to all equations, figures, tables and reset the counter %%%%%%%%%%

\begin{bibunit}[unsrt]
  
%% \section{Flowchart of the Monte Carlo Scheme}

%% \begin{figure}[h!]
%%   \centering
%%   \includegraphics[width=\linewidth]{MC_Tikz.png}
%%   \caption{A flowchart of different steps involved in simulating a GB
%%     using the proposed Monte-Carlo simulation technique.}
%%   \label{fig:fc}
%% \end{figure}

% \clearpage
% \newpage

\section{Error in the Five-Parameter GB Space}
\label{sec:Serr5}

In Figures \ref{fig:alfzs}, \ref{fig:nifzs} and \ref{fig:fefzs}, the
\% Error is plotted for all the GBs simulated for Aluminum, Nickel
and $\alpha-$Iron, respectively. The complete crystallography of the GB
is provided by specifying the GB in the boundary-plane fundamental
zone for each misorientation. Please refer to
\cite{patala2013symmetries} for further details on GB
representation.

\begin{landscape}

  \begin{figure}[!hbt]
  \centering
  \includegraphics[width=\linewidth]{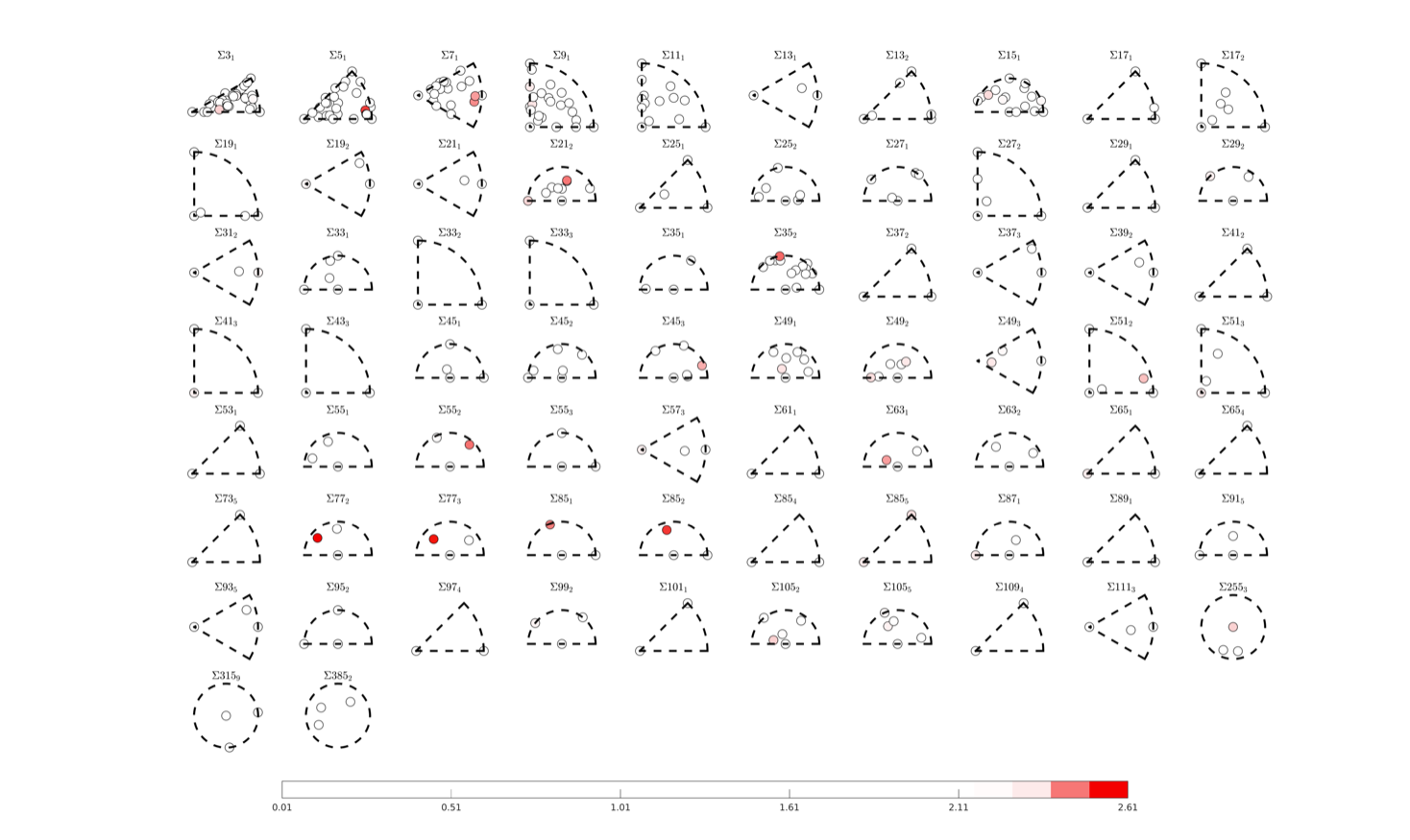}
  \caption{An illustration of the level of error for Aluminum
    simulations based on the location of the stereographic projection
    of the boundary plane of the GB on the fundamental zones of each
    $\Sigma$ rotation that has been included in the data-set.}
  \label{fig:alfzs}
\end{figure}

\begin{figure}[!hbt]
  \centering
  \includegraphics[width=\linewidth]{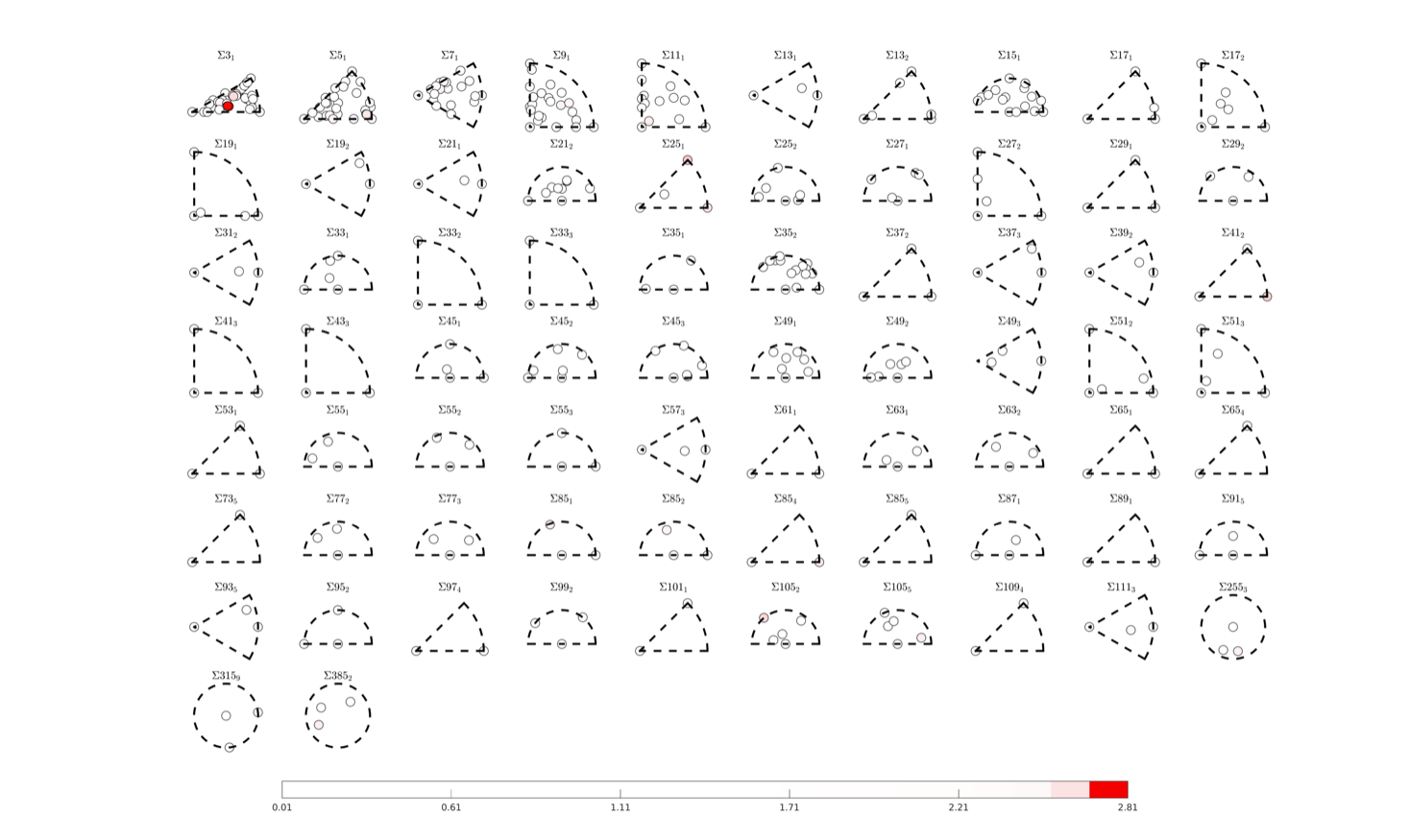}
  \caption{An illustration of the level of error for Nickel
    simulations based on the location of the stereographic projection
    of the boundary plane of the GB on the fundamental zones of each
    $\Sigma$ rotation that has been included in the data-set.}
  \label{fig:nifzs}
\end{figure}

\begin{figure}[!hbt]
  \centering
  \includegraphics[width=\linewidth]{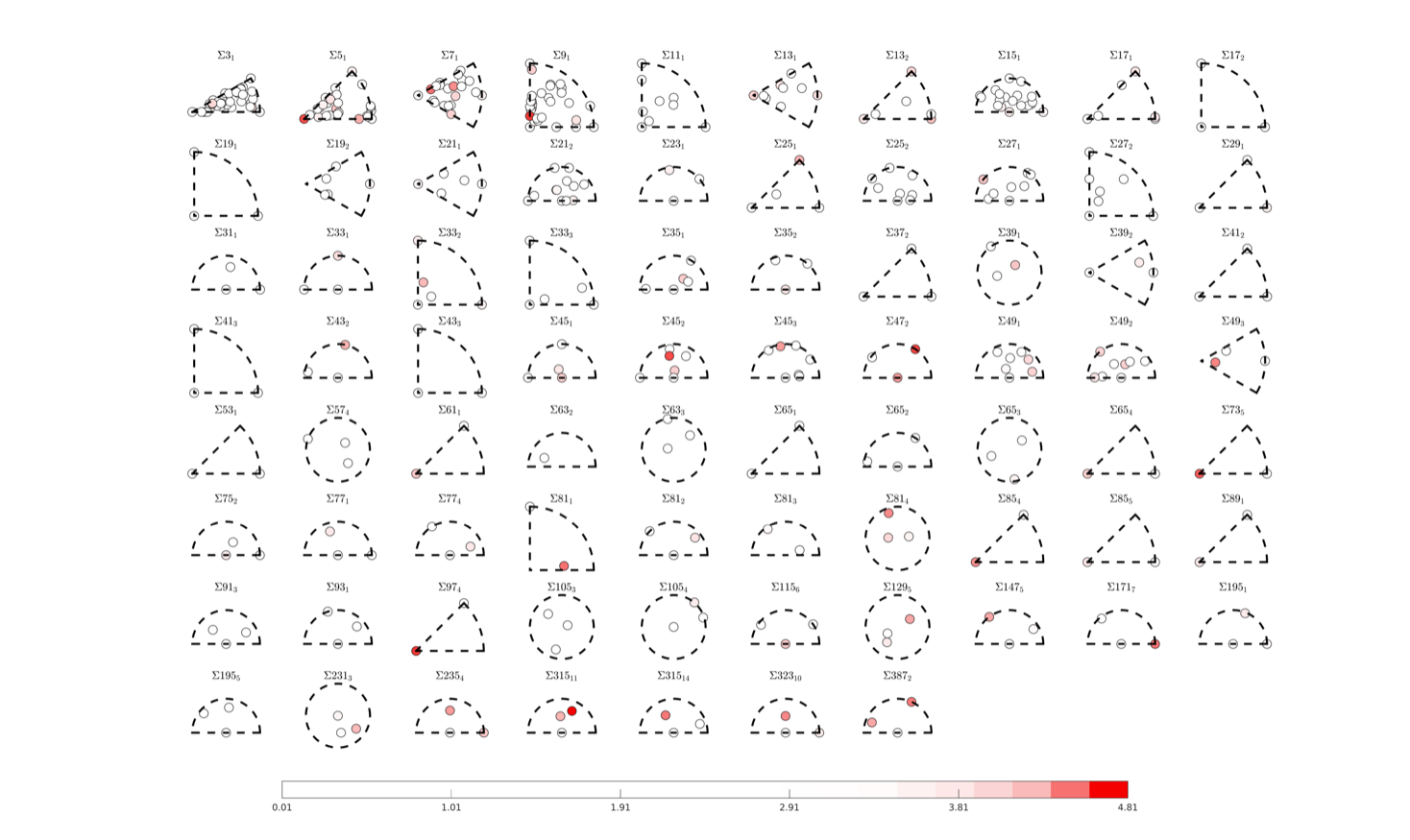}
  \caption{An illustration of the level of error for $\alpha$-Iron
    simulations based on the location of the stereographic projection
    of the boundary plane of the GB on the fundamental zones of each
    $\Sigma$ rotation that has been included in the data-set.}
  \label{fig:fefzs}
\end{figure}

\end{landscape}

\clearpage
\newpage

% \begin{figure}[h!]
%   \centering
%     \includegraphics[width=1.0\linewidth]{al_rand_mc_5000_bf_hist.png}
%   \caption{A histogram of the Error (\%), defined as the difference
%     between the minimum energies obtained through the Monte Carlo
%     scheme and the brute force algorithm (Equation~\ref{eq:error}),
%     for the 388 GBs of Aluminum, is shown. The initial configuration
%     for the MC simulation is created using a random set of microscopic
%     degrees of freedom of the GBs.}
%   \label{fig:err0}
% \end{figure}

% \section{Role of Initial GB Structure in MC Simulations}
% \label{sec:init}

% \begin{figure}[h!]
%   \centering
%   %\includegraphics[width=0.8\linewidth]{al_comp_5000_bf_hist.png}
%   \caption{\colb{Histograms of the Error (\%) for the 388 GBs of
%       Aluminum, are shown for different initial configurations.  In
%       (a), the initial configurations for the MC simulations are
%       constructed with random microscopic DOF. In (b) the maximum
%       energy and the most frequent structures (other than the minimum
%       energy structure) are chosen for initial configurations. These
%       structures are obtained from the brute-force simulations
%       performed for Aluminum GBs.}}
%   \label{fig:err1}
% \end{figure}

% \clearpage
% \newpage

\section{Evolution of GB Energy in the Monte Carlo Scheme}
\label{sec:Smcevol}

The Monte Carlo perturbations (corresponding to either the removal or
the insertion of atoms) change the structure enough that, we believe,
the GB jumps from one local minima to another. For example, shown in
figures \ref{fig:mcevol3} to \ref{fig:mcevol2} are three different
cases that reflect the evolution of energy as a function of
Monte-Carlo steps. These simulations correspond to the Aluminum GBs
with the maximum-energy structure as the initial configuration. In
general, the energy fluctuates a lot with each perturbation
corresponding to atom removal or insertion. Qualitatively, we
classified the trends in energy fluctuations, for the 388 GBs of
Aluminum, into the following categories:

\begin{enumerate}
\item GBs where the lowest energy state is visited reasonably
  frequently.  These GBs visited the lowest energy structure about 5
  to 500 times during the MC simulation. This seems to be the most
  likely case for the evolution of GB energies, i.e. there are about
  267 GBs ($\sim 69\%$) that show this behavior. Three examples are
  shown in figure \ref{fig:mcevol3}.
  
\item The cases where the lowest energy state is visited very
  frequently. These GBs visited the lowest energy structure more than
  500 times during the MC simulation. There are about 26 GBs
  ($\sim 7\%$) that show this behavior. Three examples of such
  evolution are shown in figure \ref{fig:mcevol1}.

\item GBs where the lowest energy state is visited very infrequently.
  These GBs visited the lowest energy structure less than 5 times
  during the MC simulation. There are about 95 GBs ($\sim 24\%$) that
  show this behavior. Three examples of such behavior are shown in
  figure \ref{fig:mcevol2}.
\end{enumerate}

We also show a few cases where a large increase in the energy is
observed during the MC simulation (figure \ref{fig:mcevol4}). This is
not unexpected in MC simulations because of a finite (albeit very low)
probability of acceptance of perturbations that lead to a large
increase in energy.

Finally, in figures \ref{fig:mccomp1}, \ref{fig:mccomp2},
\ref{fig:mccomp3}, and \ref{fig:mccomp4}, we plotted the energy
evolution during MC simulations with distinct initial configurations
(maximum-energy, most-frequent and random), for
$\Sigma 21 (7 \, 5 \, \bar{4})$, $\Sigma 21(\bar{1} \, 0 \, \bar{3})$,
$\Sigma 35(8 \, \bar{5} \, \bar{11})$, and
$\Sigma 5 (5 \, \bar{11} \, 8)$ GBs,
% $\Sigma 21 (7 \, 5 \, \bar{4}) (\bar{3} \, 0 \, 1)$,
% $\Sigma 21(\bar{1} \, 0 \, \bar{3})(1 \, 5 \, 8)$,
% $\Sigma 35(8 \, \bar{5} \, \bar{11})(\bar{8} \, 11 \, 5)$, and
% $\Sigma 5 (5 \, \bar{11} \, 8)(\bar{5} \, 4 \, \bar{13})$ GBs,
respectively. As such there is no discernible difference in the
evolution of energies after a few initial MC trial moves. The specific
GBs are chosen to show a diversity in the minimum energy
configurations obtained.
\begin{itemize}
\item The maximum-energy initial configuration resulted in the lowest
  energy structure for $\Sigma 21 (7 \, 5 \, \bar{4})$ GB (Figure
  \ref{fig:mccomp1})
\item For $\Sigma 21(\bar{1} \, 0 \, \bar{3})$ GB, (Figure
  \ref{fig:mccomp2}), the random configuration gave the lowest energy
  structure.
\item For $\Sigma 35(8 \, \bar{5} \, \bar{11})$ and
  $\Sigma 5 (5 \, \bar{11} \, 8)$ GBs (Figures \ref{fig:mccomp3} and
  \ref{fig:mccomp4}), starting with the most-frequent configuration
  gave the lowest energy structure.
\end{itemize}

\clearpage
\newpage

% $\Sigma 3 (1\,1\,1)(\bar{1}\,\bar{1}\,\bar{1})$
% $\Sigma 27 (1\,\bar{1}\,5)(1\,\bar{1}\,\bar{5})$
% $\Sigma 101 (1\,0\,0)(\bar{1}\,0\,0)$
% $\Sigma 29(0\,\bar{2}\,5)(0\,\bar{2}\,\bar{5})$
% $\Sigma 105(3\,1\,5)(\bar{1}\,\bar{3}\,\bar{5})$
% $\Sigma 63(\bar{1}\,10\,\bar{5})(\bar{1}\,\bar{5}\,10)$
% $\Sigma 111(11\,\bar{10}\,\bar{1})(\bar{10}\,11\,\bar{1})$
% $\Sigma 57(3\,\bar{5}\,2)(\bar{2}\,5\,\bar{3})$
% $\Sigma 93(5\,\bar{6}\,1)(\bar{1}\,6\,\bar{5})$
% $\Sigma 111(1\,1\,1)(\bar{1}\,\bar{1}\,\bar{1})$
% $\Sigma 55(\bar{3}\,\bar{1}\,0)(3\,1\,0)$
% $\Sigma 99(\bar{4}\,1\,1)(4\,\bar{1}\,1)$
% $\Sigma 95(0\,\bar{1}\,3)(1\,0\,\bar{3})$
% $\Sigma 77(\bar{9}\,5\,2)(9\,\bar{5}\,2)$
% $\Sigma 109(1\,0\,0)(\bar{1}\,0\,0)$

\begin{landscape}

  \begin{figure}
  % \centering
  \begin{subfigure}{1.0\linewidth}
    \centering
    \includegraphics[width=1.0\linewidth]{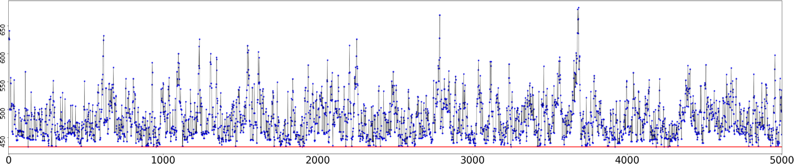}
    \caption{$\Sigma 41 (0 \, \bar{1} \, 9) (0 \, \bar{1} \, \bar{9})$}
    \includegraphics[width=1.0\linewidth]{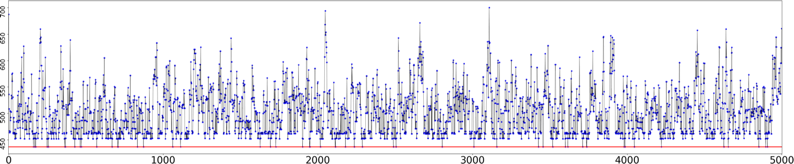}
    \caption{$\Sigma 25 (5 \, \bar{4} \, \bar{3}) (\bar{4} \, 5 \, \bar{3})$}
    \includegraphics[width=1.0\linewidth]{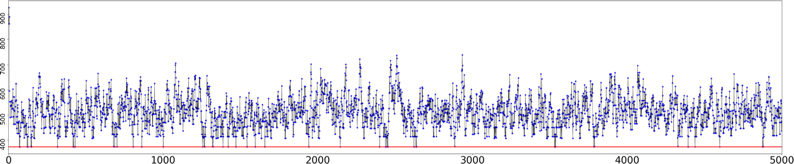}
    \caption{$\Sigma 9  (1 \, \bar{1} \, 1) (\bar{1} \, 1 \, \bar{5})$}
  \end{subfigure}%
  \caption{The low energy state is visited with reasonable
    frequency. This seems to be the most likely case.}
  \label{fig:mcevol3}
\end{figure}

\begin{figure}
  % \centering
  \begin{subfigure}{1.0\linewidth}
    \centering
    \includegraphics[width=1.0\linewidth]{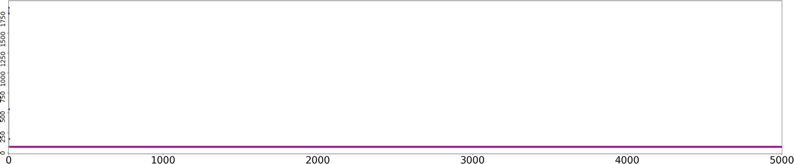}
    \caption{$\Sigma 3 \left( 1 \, 1 \, 1 \right)\left( \bar{1} \, \bar{1} \, \bar{1} \right) $}
    \includegraphics[width=1.0\linewidth]{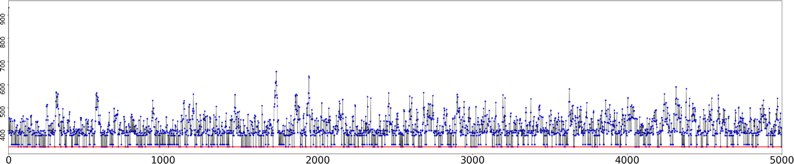}
    \caption{$\Sigma 5 \left( 10 \, \bar{1} \, 3 \right)\left( \bar{10} \, \bar{1} \, \bar{3} \right) $}
    \includegraphics[width=1.0\linewidth]{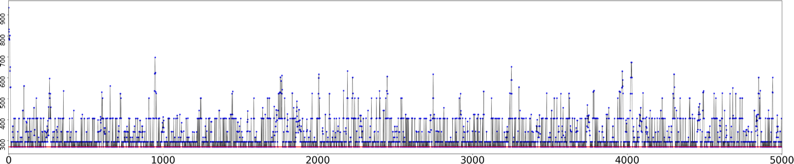}
    \caption{$\Sigma 21 \left( 1 \, 1 \, 1 \right)\left( \bar{1} \, \bar{1} \, \bar{1} \right) $}
  \end{subfigure}%
  \caption{The lowest energy structure is visited very frequently.}
  \label{fig:mcevol1}
\end{figure}

\begin{figure}
  % \centering
  \begin{subfigure}{1.0\linewidth}
    \centering
    \includegraphics[width=1.0\linewidth]{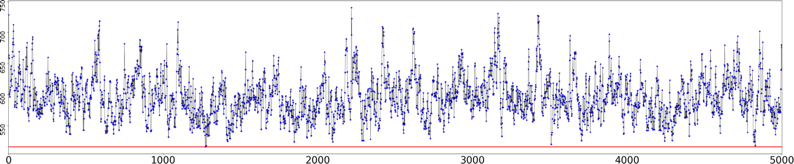}
    \caption{$\Sigma 15 (\bar{2} \, \bar{1} \, 0) (2 \, 1 \, 0)$}
    \includegraphics[width=1.0\linewidth]{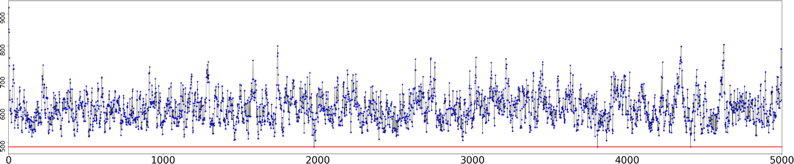}
    \caption{$\Sigma 17 (0 \, \bar{1} \, 4) (0 \, \bar{1} \, \bar{4})$}
    \includegraphics[width=1.0\linewidth]{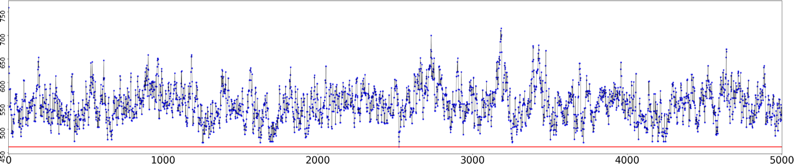}
    \caption{$\Sigma 9 (8 \, 1 \, 5) (\bar{5} \, \bar{4} \, \bar{7})$}
  \end{subfigure}%
  \caption{The lowest energy structure is rarely visited.}
  \label{fig:mcevol2}
\end{figure}

\begin{figure}
  % \centering
  \begin{subfigure}{1.0\linewidth}
    \centering
    \includegraphics[width=1.0\linewidth]{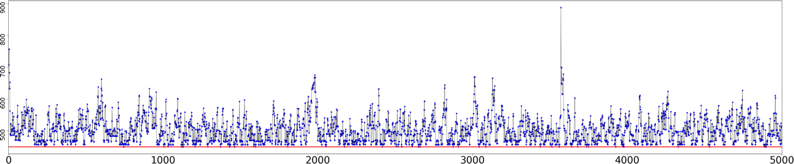}
    \caption{$\Sigma 21 (5 \, \bar{4} \, \bar{1}) (\bar{4} \, 5 \, \bar{1})$}
    \includegraphics[width=1.0\linewidth]{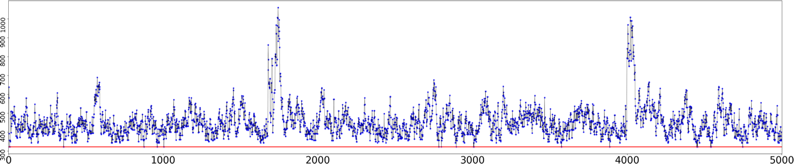}
    \caption{$\Sigma 17 (1 \, 5 \, 5) (\bar{1} \, \bar{7} \, \bar{1})$}
    \includegraphics[width=1.0\linewidth]{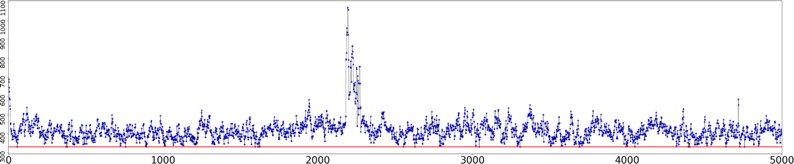}
    \caption{$\Sigma 3  (3 \, 1 \, 0) (\bar{8} \, 1 \, \bar{5})$}
  \end{subfigure}%
  \caption{A few cases where the local perturbation results in a large
    increase in energy. Such a behavior is not unexpected in Monte
    Carlo simulations.}
  \label{fig:mcevol4}
\end{figure}

\begin{figure}
  % \centering
  \begin{subfigure}{1.0\linewidth}
    \centering
    \includegraphics[width=1.0\linewidth]{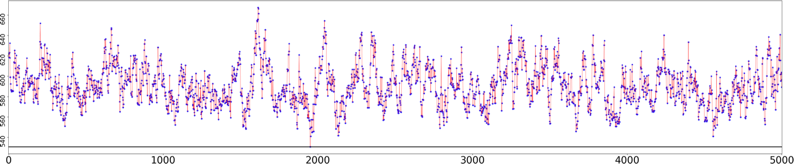}
    \caption{Maximum Energy Structure; minimum energy found = \textbf{530.7} mJ/m$^2$}
    \includegraphics[width=1.0\linewidth]{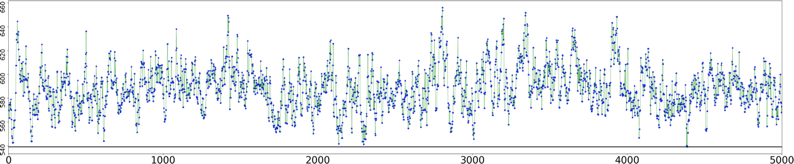}
    \caption{Most Frequent Structure; minimum energy found = 539.1 mJ/m$^2$}
    \includegraphics[width=1.0\linewidth]{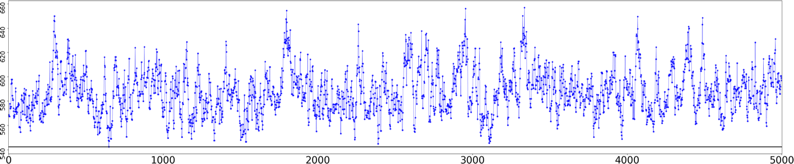}
    \caption{Random Structure; minimum energy found = 542.4 mJ/m$^2$}
  \end{subfigure}%
  \caption{The evolution of energies during the MC simulation for
    three initial configurations (a) the maximum energy structure, (b)
    the most frequent structure and (c) the random structure, for
    $\Sigma 21 (7 \, 5 \, \bar{4}) (\bar{3} \, 0 \, 1)$. The minimum
    energies found in each simulation are provided in the captions.}
  \label{fig:mccomp1}
\end{figure}

\begin{figure}
  % \centering
  \begin{subfigure}{1.0\linewidth}
    \centering
    \includegraphics[width=1.0\linewidth]{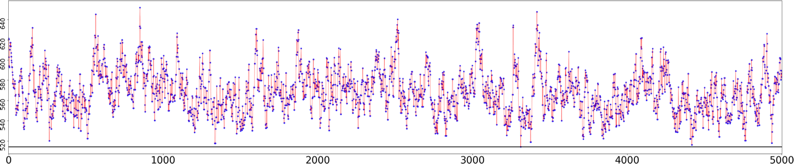}    
    \caption{Maximum Energy Structure; minimum energy found = 514.3 mJ/m$^2$}
    \includegraphics[width=1.0\linewidth]{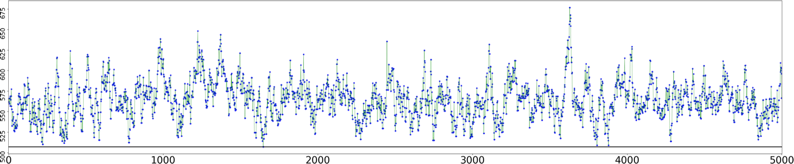}
    \caption{Most Frequent Structure; minimum energy found = 507.1 mJ/m$^2$}
    \includegraphics[width=1.0\linewidth]{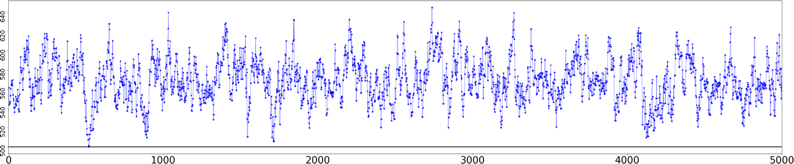}
    \caption{Random Structure; minimum energy found = \textbf{500.1} mJ/m$^2$}
  \end{subfigure}%
  \caption{The evolution of energies during the MC simulation for
    three initial configurations (a) the maximum energy structure, (b)
    the most frequent structure and (c) the random structure, for
    $\Sigma 21(\bar{1} \, 0 \, \bar{3})(1 \, 5 \, 8)$. The minimum
    energies found in each simulation are provided in the captions.}
  \label{fig:mccomp2}
\end{figure}

\begin{figure}
  % \centering
  \begin{subfigure}{1.0\linewidth}
    \centering
    \includegraphics[width=1.0\linewidth]{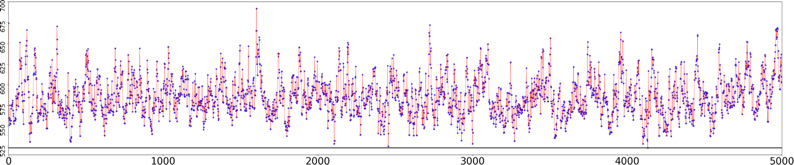}
    \caption{Maximum Energy Structure; minimum energy found = 524.0 mJ/m$^2$}
    \includegraphics[width=1.0\linewidth]{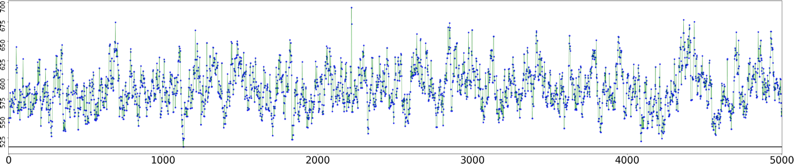}
    \caption{Most Frequent Structure; minimum energy found = \textbf{513.3} mJ/m$^2$}
    \includegraphics[width=1.0\linewidth]{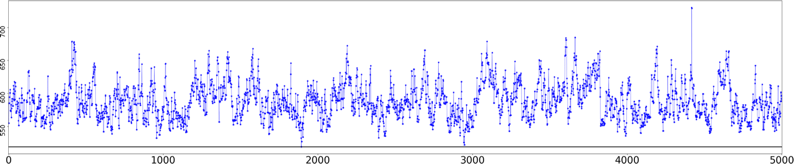}
    \caption{Random Structure; minimum energy found = 519.0 mJ/m$^2$}
  \end{subfigure}%
  \caption{The evolution of energies during the MC simulation for
    three initial configurations (a) the maximum energy structure, (b)
    the most frequent structure and (c) the random structure, for
    $\Sigma 35(8 \, \bar{5} \, \bar{11})(\bar{8} \, 11 \, 5)$. The minimum
    energies found in each simulation are provided in the captions.}
  \label{fig:mccomp3}
\end{figure}

\begin{figure}
  % \centering
  \begin{subfigure}{1.0\linewidth}
    \centering
    \includegraphics[width=1.0\linewidth]{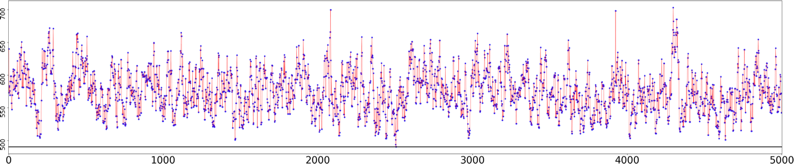}
    \caption{Maximum Energy Structure; minimum energy found = 490.6 mJ/m$^2$}
    \includegraphics[width=1.0\linewidth]{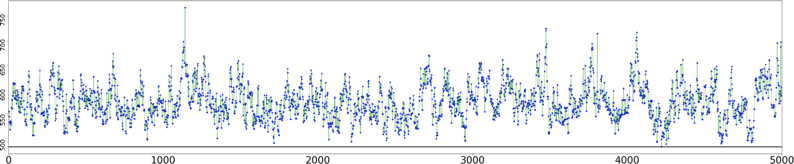}
    \caption{Most Frequent Structure; minimum energy found = \textbf{488.4} mJ/m$^2$}
    \includegraphics[width=1.0\linewidth]{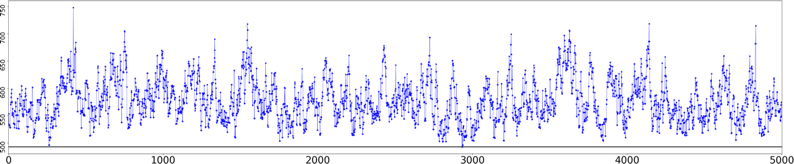}
    \caption{Random Structure; minimum energy found = 493.3 mJ/m$^2$}
  \end{subfigure}%
  \caption{The evolution of energies during the MC simulation for
    three initial configurations (a) the maximum energy structure, (b)
    the most frequent structure and (c) the random structure, for
    $\Sigma 5 (5 \, \bar{11} \, 8)(\bar{5} \, 4 \, \bar{13})$. The minimum
    energies found in each simulation are provided in the captions.}
  \label{fig:mccomp4}
\end{figure}

\end{landscape}
% \putbib

% \end{bibunit}

\section*{Supplemental References}
\putbib[main]
\end{bibunit}

% \section*{References}
% % `Elsevier LaTeX' style
% \bibliography{supp}

\end{document}